\documentclass[acmsmall,screen]{acmart}

\AtBeginDocument{%
  }

\setcopyright{acmlicensed}
\copyrightyear{2025}
\acmYear{2025}
\acmDOI{XXXXXXX.XXXXXXX}

\usepackage{graphicx}
\usepackage[ruled]{algorithm}
\usepackage{hyperref}
\usepackage{algorithmicx}
\usepackage{algpseudocode}
\usepackage{booktabs}
\usepackage{xcolor}
\usepackage{multirow}
\usepackage{booktabs}
\usepackage{tcolorbox}
\usepackage{subcaption}
\usepackage{longtable}
\usepackage{xspace}
\usepackage{tikz}
\usetikzlibrary{positioning}
\newcommand\etal[0]{\emph{et al.}\xspace}

\newcommand{\eg}{\emph{e.g.,}\xspace}

\newcommand{\hide}[1]{}
\newcommand{\prefix}[1]{\MakeUppercase{#1}}
\newcommand{\prefixours}{BISS\xspace}

\newcommand{\variants}{\mathcal{A}}
\newcommand{\tests}{\mathcal{T}}

\newcommand{\solves}{\text{\new{findWeightsAndCheck}}}

\definecolor{palblue}{HTML}{1E88E5}
\definecolor{palyel}{HTML}{FFC107}
\definecolor{palgreen}{HTML}{3A9502}
\definecolor{red}{HTML}{FF4433}

\colorlet{todocolor}{red}

\newcommand{\new}[1]{#1}

\newtheorem{definition}{Definition}[section]

\begin{document}

\title{Efficiently Ranking Software Variants with Minimal Benchmarks}




\author{Théo Matricon}
\email{theo.matricon@inria.fr}
\orcid{0000-0002-5043-3221}
\affiliation{%
\institution{Univ. Rennes, Inria, CNRS, IRISA}
\city{Rennes}
\country{France}
}

\author{Mathieu Acher}
\email{mathieu.acher@inria.fr}
\orcid{0000-0003-1483-3858}
\affiliation{%
\institution{Univ. Rennes, Inria, CNRS, IRISA, IUF}
\city{Rennes}
\country{France}
}

\author{Helge Spieker}
\affiliation{%
  \institution{Simula Research Laboratory}
  \city{Oslo}
  \country{Norway}}
\email{helge@simula.no}
\orcid{0000-0003-2494-4279}

\author{Arnaud Gotlieb}
\affiliation{%
  \institution{Simula Research Laboratory}
  \city{Oslo}
  \country{Norway}}
  \email{arnaud@simula.no}
  \orcid{0000-0002-8980-7585}

\renewcommand{\shortauthors}{Matricon et al.}

\begin{abstract}
Benchmarking is a common practice in software engineering to assess the qualities and performance of software variants,
coming from multiple competing systems or from configurations of the same system.
 Benchmarks are used notably to compare and understand variant performance, fine-tune software, detect regressions, or design new software systems.
 The execution of benchmarks to get a complete picture of software variants is highly costly in terms of computational resources and time.
In this paper, we propose a novel approach for reducing benchmarks while maintaining stable rankings, using test suite optimization techniques. That is, we remove instances from the benchmarks while trying to keep the same rankings of the variants on all tests.
 Our method, BISection Sampling, \prefixours, strategically retains the most critical tests and applies a novel divide-and-conquer approach 
 to efficiently sample among relevant remaining tests. We experiment with datasets and use cases from LLM leaderboards, SAT competitions, and configurable systems for performance modeling. Our results show that our method outperforms baselines even when operating on a subset of variants. Using \prefixours, we reduce the computational cost of the benchmarks on average to 44\% and on more than half the benchmarks by up to 99\% without loss in ranking stability.
\end{abstract}



\received{20 February 2007}
\received[revised]{12 March 2009}
\received[accepted]{5 June 2009}

\maketitle

    \section{Introduction}

\begin{figure*}[ht!]
    \centering
    \includegraphics[width=\linewidth]{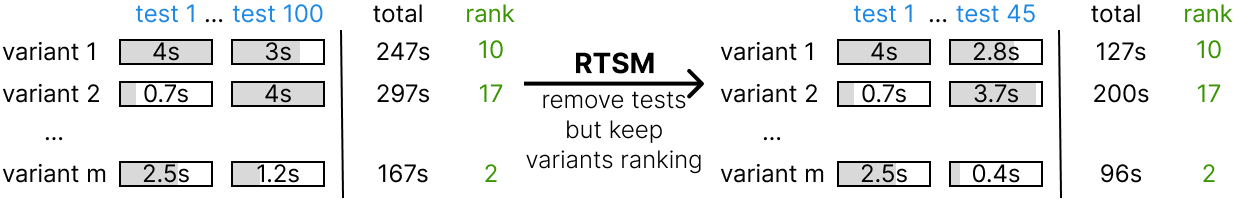}
    \caption{High Level Overview of Ranked Test Suite Minimization (RTSM) which aims at minimizing benchmarks while keeping the ranking of the variants corrects enabling reliable, fast and cheap evaluation of new variants.}
    \label{fig:big-picture}
    \Description{The figure shows a set of variants and 100 tests being reduced by solving the RTSM problem to 45 tests while producing the same overall ranking of the variants.}
\end{figure*}
Benchmarking is a fundamental practice in software engineering that focuses on evaluating the qualities and performance of software systems.
It is also essential in computer science for tasks such as optimization of algorithms or evaluation of system performance~\cite{bartzbeielstein2020benchmarkingoptimizationbestpractice,Olson2017,Kerschke2019, Malan2013, Munoz2015,benchopt}.
In computational science, scientists use benchmarking to explore, compare and validate the performance of complex variants of simulations and models, facilitating advances in research and development~\cite{Donoho2024Data,Guyon2020}.
 Benchmarks can be seen as a special kind of test that is used to measure software performance such as execution time, energy consumption, or accuracy.
Unlike unit or integration tests that return binary or ternary test results (i.e., Pass/Fail/inconclusive), benchmarking returns numerical evaluations, which makes them more challenging to work with.
This is why software system benchmarking usually involves comparisons with other software variants.
Hence, benchmarks are executed on different variants, whether they originate from different competing, alternative systems, or various configurations of the same system.

These benchmarks, 
workloads, test sets, performance tests, inputs, or problems
typically consist of a set of test cases to measure and compare the software variants' performance.
 The ranking of software variants based on benchmark results provides an objective quantitative measure of their relative performance.
This ranking not only highlights discrepancies in variant performance, but also offers a concrete basis for comparing the diversity of software solutions across different problem domains.
In addition, rankings can reveal patterns and trends in performance across various types of inputs or problem characteristics, providing valuable, data-driven insights for software optimization and development strategies.

Benchmarks and performance comparisons are crucial for understanding the performance characteristics of software variants, configuring and fine-tuning software, detecting regressions, and guiding the design of new systems.
 However, executing comprehensive benchmarks to achieve a complete understanding of software variants can be extremely costly in terms of computational resources and time.
For instance, LLM leaderboards~\cite{liu2023codegeneratedchatgptreally} or SAT competitions~\cite{froleyks2021sat, heule2019sat} are two well-known examples of benchmarking competitions that require extensive computational resources to evaluate the performance of variants.
 The need for more efficient methods to maintain accurate rankings of software variants while minimizing the computational load is evident in several concrete scenarios:
\begin{itemize}
\item \textbf{Comparison:} In competitive settings or for the sake of improvements, numerous software variants must be evaluated and ranked.
For instance, leaderboards are becoming more and more available to provide engineers with a variety of benchmarks and the associated performance of the variants.
Full benchmarks are often prohibitively expensive, especially with many entries.
A minimized benchmark significantly reduces computational costs while still accurately ranking top performers and revealing their strengths and weaknesses.
\item \textbf{Regression testing}: Executing reduced benchmarks on each commit or release allows for frequent performance checks,
enabling quick detection of regressions' rankings, for example, in continuous integration. 
This approach facilitates rapid iterations and more thorough exploration of alternatives in both software engineering and scientific computing, with better-informed decisions and continuous performance assessment; 
\item \textbf{Performance modeling, extrapolation, and configuration tuning:} Engineers often use mathematical models and simulations to predict software performance under various conditions or to find optimal configurations~\cite{balsamo04modelbased}. By selecting a subset of representative benchmarks, they can efficiently train models that capture essential performance distributions. When benchmarking is training~\cite{bartzbeielstein2020benchmarkingoptimizationbestpractice}, this process reduces the need for exhaustive execution, allowing engineers to efficiently identify performance bottlenecks and optimize system configurations based on workload sensitivity (\eg as recently shown in~\cite{muhlbauer2023,lesoil2023}). 
\item \textbf{Benchmark design and elaboration:} In case benchmarks are repetitive or identical, a strategy to design better benchmarks can be to focus on the most informative tests and extend them;


\item \textbf{Benchmarking a new system:} When new systems (variants) are considered, reduced benchmarks can be leveraged at a better cost to get a preliminary performance comparison with competitors.
\end{itemize}

In short, our goal is to equip software engineers, computational scientists, and researchers
with an objective method to reduce the computational cost of benchmarks while still effectively assessing competing software-based systems (see \autoref{fig:big-picture}).
It can boost innovation and productivity by enabling more efficient evaluation or simply reducing the cost of benchmarking.

However, all of these scenarios also highlight the potential dangers of reduced benchmarks that do not fully represent the entire set of benchmarks.
 To our knowledge, no previous work has addressed the problem of minimizing benchmarks while preserving the ranking of multiple software variants (see also more details in~\autoref{sec:related_work}).
Previous research has extensively explored benchmarking and testing~\cite{harman2012RT}, focusing on optimization, prioritization, and reduction of the test suite, but primarily for single systems, functional tests, and not for competing software variants.
 The works on software product lines (SPL) focus on the analysis of multiple products~\cite{10.1145/3382025.3414967, Al-Hajjaji2017, Elbaum2014, Ensan2011, Kumar2016, Lachmann2017, Lachmann2015, Lity2019, Lochau2012, Marijan2019, Marijan2013, Marijan2017, Marijan2017titan, PradoLima2020, Runeson2012, wang_cost-effective_2015, Wang2016, Xu2013},
 but assume a single code base and seek to minimize the number of test cases while usually maximizing the functional coverage of features or code (not performance).

Our optimization goal thus differs and consists in preserving the performance ranking of potentially independent software variants. 
 In this paper, we propose a novel method to reduce the computational cost of benchmarks while still effectively ranking software variants using test suite optimization techniques.
 \new{Drawing from the fact that the problem is NP-hard, we turn to heuristics with anytime algorithms because even small reductions can provide benefits; there is no need for exact optimality most of the time.}
Our method, called \prefixours for BISection Sampling, is based on the fact that some tests cannot be removed and form the essential core of the benchmark. Furthermore, once some tests are removed, other tests become crucial as well.
This enables us to use a bisection sampling routine to sample among relevant tests cleverly. We couple it with a divide-and-conquer algorithm to reduce the variance and speed up the search, and with an automatic restart (iterative solving) as long as we find improvements. Our contributions include:

\begin{itemize}
\item The introduction of a new problem in Software Testing research, that we frame as {\it the ranked test set minimization problem (RTSM)};
\item A novel divide-and-conquer method exploiting a prediction model and parameterized search to practically solve the RTSM problem and effectively prune test suites;
\item Experimental results showcasing, on 50 scenarios and more than 30 different systems, that our approach outperforms multiple baselines and its impacts, enabling the removal of 20\% up to 99\% of tests depending on the accuracy requested;
\item An artifact with all the datasets and benchmarks needed to reproduce the results.
\end{itemize}



The remainder of this paper is organized as follows: \autoref{sec:related_work} discusses related work further. In~\autoref{sec:definition} we introduce the problem. In~\autoref{sec:technical} we first describe our method and then in \autoref{sec:baselines} propose baselines to compare to, including implementation details. We provide experimental results in \autoref{sec:experiments}. Then we focus on selected case studies in ~\autoref{sec:case_studies}. We discuss threats to validity in~\autoref{sec:threats}. 
\new{We draw conclusions, provide recommendations for developers and benchmark designers and users, and discuss future work in~\autoref{sec:impact}.}


    \section{Related work}
\label{sec:related_work}


Numerous work on regression testing and test suite minimization, selection, and prioritization has been published~\cite{harman2012RT}.
The principle is to improve the test suite of a single software system, typically to reduce computational cost as test suites tend to grow in size.
 Test suite minimization seeks to prune test suites from redundant test cases to reduce the number of tests to run.
We pursue a related, yet different, goal with two key differences: we are interested in performance of software and we consider multiple software variants.


Research in software performance engineering explores benchmarking at different levels (\eg system-level tests and method/statement-level tests) and with different approaches (\eg blackbox vs whitebox)~\cite{microbenchmarks2020,weyuker2000experience, jian2015survey, menasce2002load, blot2022comprehensivesurveybenchmarksautomated, nguyen2014industrial, alghamdi2020towards}.
For example, there are techniques for microbenchmarking that measure the performance of \eg single functions~\cite{microbenchmarks2020,Maricq2018, microbenchmarks2020,costa2019whats}. 
Ding \etal \cite{ding2020towards} studied how unit tests can be used to assess performance properties.
All of these works do not aim to preserve the performance ranking of multiple variants.

In close proximity to our work are contributions related to software product lines (SPLs) and configurable systems~\cite{thum2014classification, rhein2018variability, 10.1145/3382025.3414967, Al-Hajjaji2017, Elbaum2014, Ensan2011, Kumar2016, Lachmann2017, Lachmann2015, Lity2019, Lochau2012, Marijan2019, Marijan2013, Marijan2017, Marijan2017titan, PradoLima2020, Runeson2012, wang_cost-effective_2015, Wang2016, Xu2013, sway}.
In this line of research, the goal is to consider a set of related products or configurations coming from a common code base (an SPL or a configurable system)
and to optimize testing activities for these products.
 Existing SPL works explore the three
traditional regression testing techniques for single software systems~\cite{harman2012RT}, namely minimization~\cite{wang_cost-effective_2015},
selection~\cite{Lity2019, Lochau2012, Wang2016, Xu2013}, and prioritization \cite{Al-Hajjaji2017, Lachmann2017, Lachmann2015} of test cases.
Wang \etal propose to optimize test suites considering effectiveness measures such as test minimization percentage, pairwise coverage, fault detection capability~\cite{wang_cost-effective_2015} but these works ignore software performance properties and ranking information.
Mendonça \etal introduce FeaTestSelPrio a feature-oriented test case selection and prioritization approach that improves failure detection
and reduces test execution time by linking test cases to feature implementations and prioritizing based on feature coverage~\cite{mendonca_feature-oriented_2024}.
Our work differs from several perspectives: (1) we target variants from multiple code bases;
(2) we leave apart variant features: such features can be unavailable in our context, where we only have access to the software variants themselves;
(3) performance calls to consider quantitative properties of software, whereas test suites are typically qualitative (e.g., Fail/Pass or Inconclusive);
(4) our optimization goal is to preserve the (performance) ranking of software variants, while prior works consider failures or feature/code coverage.


Instead of reducing benchmarks, a more common problem is the selection of relevant software configurations for a given problem. 
Several configuration sampling strategies have been proposed in the literature
on SPLs and configurable systems, either to find and cover as
many faults as possible~\cite{han2016empirical, thum2014classification}, or to measure and learn the performance of the configuration space~\cite{alvespereira:hal-02148791,alvespereira:hal-02356290,kaltenecker2020interplay}.
Their focus is on the selection of configurations (variants) to test, rather than on the selection of test cases to run.
Recent empirical findings from Mühlbauer \etal~\cite{muhlbauer2023} and Lesoil \etal~\cite{lesoil2023} show that program inputs (or workloads), typically part of a benchmark, and configuration options can result in substantial performance variations. The sheer volume of configurations and benchmark elements creates high computational costs, making it impractical to learn a performance model for each test case. These performance variations can alter the ranking of software variants, highlighting the need for careful benchmark selection to ensure accurate assessments.
Nair \etal~\cite{nair2017} tackle configuration optimization through rank-based selection assuming a given workload and relying on potentially imprecise performance models. In contrast, our approach focuses on reducing the benchmark suite itself to preserve the ranking across software variants, rather than identifying optimal configurations. Existing work, such as Nair \etal, is not applicable to benchmark reduction, as it focuses on optimizing configurations within a single workload and does not address the challenge of selecting benchmarks to maintain ranking consistency across multiple variants.

Multimorphic testing evaluates the coverage of a test suite with respect to a quantitative property by generating multiple software variants to reveal performance differences \cite{multimorphic}. 
\new{It assesses} performance test suite effectiveness by identifying tests that can "kill" under-performing configurations and calculating a dispersion score to measure coverage across variants. While we operate in a similar context, our focus differs fundamentally: instead of the dispersion score, we aim to ensure that the minimized benchmark set preserves the correct performance ranking among competing variants, which is essential for various scenarios described above.
The automated algorithm selection problem is subject to intensive research~\cite{Kerschke2019, hutter2011, Hutter2014, xu2008, thornton2013} with many applications in artificial intelligence (such as SAT, CSP, QBF, ASP, or scheduling~\cite{Kerschke2019}).
Given a computational problem, a set of algorithms, and a specific problem instance to be solved, the challenge is to determine which of the algorithms can be selected to perform best on that instance.
For example, SATZilla uses machine learning to select the most effective algorithm from a portfolio of SAT solvers for a given SAT formula~\cite{xu2008}.
 In addition, determining which of two algorithms performs best on a benchmark, without running the whole benchmark, has been discussed in~\cite{matricon21comparison}. However, the focus is on two variants and the cost of running variants, not on benchmarks. 
 A critical factor in automated algorithm selection is the set of problem instances (benchmarks) used for \eg training models to facilitate optimal algorithm selection. 
 Our work contributes to this area by providing a software engineering methodology that can be applied to various AI problems.
\section{Problem Statement: Ranked Test Suite Minimization (RTSM)}\label{sec:definition}




In software testing, test suite minimization focuses on reducing the number of tests needed while still ensuring that all requirements are met. Traditionally, these functional tests check Boolean requirements, where a test either passes or fails, similarly to solving set-cover instances. However, many scenarios involve performance-related tests that have continuous and quantitative answers. 

\begin{definition}
    Given a set of configuration variants $\variants$, a {\bf performance value}  is a function $t: \variants \to \mathbb{R}$ that associates a performance value (e.g. estimation, measurement, execution time, memory consumption, etc.) with each variant of a configurable system.
\end{definition}

 Individual performance values are usually difficult to get and verify. For example, checking that a system runs exactly in 2.42 seconds, as it has been registered from previous executions or predicted by an ML model, is very unlikely. Comparison of this with other observations coming from other variants' performance estimations is usually needed. 
 Ranking performance estimations are a way to evaluate how a system operates and is widely used \cite{bartzbeielstein2020benchmarkingoptimizationbestpractice, lesoil2023}. 

Given a set of configuration variants, we can typically estimate their performance on a set of test inputs, sum their performance over these inputs, and rank each variant accordingly.
Intuitively, given a full benchmark obtained from each variant's performance estimation and its ranking of the variants, our goal is to compress the benchmark while preserving the ranking.

\begin{definition}\label{def:rtsm}
    The {\bf rank of a variant} $v \in \variants$ for a test suite $\tests$ is the rank $r$ of its performance estimation $perf(v, \tests)=\sum_{t \in \tests} t(v)$, precisely:
    \new{    $$r(v) = | \{v' \in \variants \ | \ perf(v, \tests) < perf(v', \tests)  \}|$$}
    The variant with the maximal performance estimation is ranked $1$ and so on.
    Note that this ranking computes a total ordering over the variants in $\variants$. \new{The above formula assumes that there is no tie. In practice, ties might occur and a tie-breaking strategy is required.}
\end{definition}
\new{In the experiments of this paper, ties are broken by ranking tied solvers alphabetically but ties never occur.}

\begin{definition}\label{def:wrtsm}
    Given a set of variants $\variants$ and a test suite $\tests$, the {\bf Ranking Test Suite Minimization problem (RTSM)} aims to find a smallest $\tests' \subset \tests$ and $W$ such that the ranking of variants $\variants$ for $\tests'$ is preserved (i.e., the same as the ranking of variants $\variants$ on $\tests$) where $W$ gives a weight to each test in $\tests'$ for ranking. Note that several subsets $\tests'$ can be solutions of RTSM.
\end{definition}

Looking at the left part of the \autoref{fig:big-picture}, we have an initial set of variants along with the \textcolor{palblue}{tests} set and the performance estimation of each pair variant test. Their \textcolor{palgreen}{ranking} is determined by classifying the variants in ascending order with respect to the summed performance estimates. A solution to the RTSM problem is given on the right part of the figure. Keeping only 45 tests as a subset of \textcolor{palblue}{tests}, the ranking of variants is preserved even though the total performance estimation has changed.   

Note that test suite minimization is an instance of RTSM when one just ignores the ranking. As test suite minimization is NP-hard \cite{goliebissta14}, we can deduce from the previous remark that RTSM is also NP-hard. \new{That means, while finding the minimal subset is NP-hard, finding a solution remains easy: $\tests$ is a solution.}
\new{In this work, we do not attempt to find an optimal solution to this NP-hard problem, but propose an empirically efficient heuristic solution, designed for practical applications.}

\begin{definition}
    Given a set of variants $\variants$ and a test suite $\tests$ with a cost associated with each test, the {\bf Weighted Ranking Test Suite Minimization problem (WRTSM)} aims to find $\tests' \subset \tests$ and $W$ such that the ranking of variants $\variants$ for $\tests'$ is preserved (i.e., the same as the ranking of variants $\variants$ on $\tests$) and that minimizes $cost(\tests')=\sum_{t \in \tests'} cost(t)$ where $W$ gives a weight to each test in $\tests'$ for ranking. 
\end{definition}

\new{We will use the weighted approach whenever we want to reduce the cost of running a test suite and not just the number of tests. This is of course also NP-hard, that is finding the test suite with minimal cost is NP-hard.}

\paragraph*{Error Metric}
Unfortunately, there is no ground truth; therefore, we will compare the rankings obtained using the solution \new{found and the original ranking prior to minimization}.
There are multiple metrics to compare the rankings, the two most commonly used ones are Kendall's~\cite{kendall} and Spearman's~\cite{spearman} rank correlation coefficient.
Both give a score between -1, the two vectors offer opposite rankings, and 1, the two vectors offer exactly the same rankings.
We chose to use Kendall's, which is directly proportional to the number of variants correctly ranked.
Choosing Spearman or Kendall does not matter when the aim is a score of 1, that is, the same ranking, because they both coincide for this data point.
Both tests are non-parametric hypothesis tests on the relations between the two rankings.
Kendall can be easily computed, assume that a pair of variants are concordant if they are ranked in the same order in both rankings and discordant otherwise, then we have:
\begin{equation*}
    \text{Kendall's } \tau = \frac{\#|concordant\ pairs| - \#|discordant\ pairs|}{\#|pairs|}
\end{equation*}
Special cases occur when there are ties, but in our scenarios, ties never occur.

\paragraph*{Multi-metric} The current formulations in \autoref{def:rtsm} and \autoref{def:wrtsm} use a single performance metric, but can be easily adapted to multiple metrics.
When multiple metrics are present, compute the Kendall for each metric independently and take the worst one.
For example, if we wanted a Kendall of .99 in the multi-metric case, then on all metrics the Kendall must at least be .99.


   \section{Benchmark minimization with \prefixours}\label{sec:technical}

\paragraph*{Novelty} As explained in \autoref{sec:related_work}, most existing methods for minimizing the test suite require prior information such as test results, historical data, etc.
\new{They are all mostly practical approaches offering no guarantees due to the scalability required. }
However, the scenarios for RTSM are very diverse, requiring handcrafted features by experts for their domain if we want to use a white-box approach, which are often not available or are too costly to acquire. Instead, we choose to look at black-box approaches.
We have no features about the variants; we have just an identifier to be able to identify them, and this is the same for the tests.
So, as input, we only have the performance matrix, that is, a matrix that contains the performance of each variant on each test.
Then techniques such as principal component analysis (PCA) or racing techniques~\cite{LOPEZIBANEZ201643} should solve this problem. We will see that we tested these techniques and they do not solve our problem.
Indeed, these classic techniques can work to some extent, but we are interested in objectives that are a loose proxy of our interest: the ranking of variants. Whereas most of these techniques are interested in more fine-grained information: the magnitude of the differences, they care if you are twice as fast or ten times as fast, whereas in RTSM it does not matter.

\paragraph{Intuition} \new{We adopt an empirical approach in which }the core idea is that we do not care about the individual performances of the tests; we only care about the relative performance of a variant to another across tests.

\autoref{def:rtsm} gives us the possibility of assigning weights to $W \new{\in \mathbb{R}^{|\tests'| \times 1}}$: Given a subset of the test set, we can use a linear regression to optimize $W$ to obtain the best coefficients.
Let $P\in\mathbb{R}^{|\variants|\times|\tests|}$ be the performance matrix of the variants on the tests, then, given a subset $\tests$' of $\tests$, we can optimize:
\begin{align*}
    \new{\min_W || P \ \mathbf{1}_{|\tests|\times 1} - P_{* \tests'} W ||^2}
\end{align*}
\new{where  $\mathbf{1}_{|\tests|\times 1} \in \mathbb{R}^{|\tests|\times 1}$ is a matrix of ones and $P_{* \tests'}$ is the matrix $P$ where only the columns concerning the tests in $\tests'$ were kept. 
This is an ordinary least squares where the goal is to predict the aggregated performance on the full set of tests $\tests$ from the set of tests that were kept $\tests'$.  }
Our goal is therefore to sample relevant subsets $\tests'$.
Formally, we assume that we have the procedure $\solves_T(T')$, \new{that} solves this linear regression and checks if the Kendall target is reached with $T'$ as a solution to the problem with the complete test set being $T$. 
If it is, then the procedure returns $true$, otherwise it returns \textit{false}. 
\new{ $\solves_{K\cup P}(K)$ has complexity O($|\variants||K|^2|P|)$, where $\variants$ is the set of variants. It is polynomial, but this is expected, as it only solves a linear regression problem and then computes if the solution is valid within the target Kendall value. 
It does not find the candidate subset $K$ nor does it prove that the solution is optimal.}

As we have explained earlier, the problem is NP-hard, and covering all possible solutions is intractable for small test suites; for a test suite of size 30, there are more than a billion possible solutions.
\new{This is why our approach does not focus on finding an optimal solution. 
Instead, the goal is to produce a solution within the constraints. 
Notice that providing a solution can be considered trivial since in all formulations the initial test suite $\tests$ is a valid solution. Our approach focuses on sampling better candidates than $\tests$.}
The process is a bit akin to dimensionality reduction; each test acts as a dimension, and we want to reduce the dimension while keeping the ranks intact. 
We will take inspiration from the Johnson–Lindenstrauss lemma~\cite{johnson1984extensions}, which states that a random projection in the ``low'' dimension has a high probability of keeping the original distances between the data points in the original high dimension. 
The dimension that the lemma gives us is logarithmically dependent on the original dimension and grows inversely proportional to the square of the minimum distance between two data points. 
In other words, the \new{more similar the performance of two variants is}, the more likely it is that \new{more tests are needed to rank them correctly}. 
Therefore, our objective is to sample subsets of tests that are likely to result in a solution.

\subsection{Bisection Sampling}\label{sec:bisect}

Obviously, we would like to sample more intelligently than pure random \new{subsets of $\tests$}.
For that, we need to improve on two points: reducing the sampling variance and eliminating tests faster than one at a time.

\paragraph*{Variance reduction} The idea is to draw on the following fact: If we add more tests to a solution to the RTSM problem, then it is also a solution. Then, conversely, if we have a set of tests that is not a solution, none of the subsets is a solution.
Given a solution $T$ for each test $t \in T$ we can check if $\solves_T(T\setminus \{t\})$ is \textit{true}, that is, if with all the information except for test $t$ we have a solution.
If $\solves_T(T\setminus \{t\})$ is \textit{false} then it must be that the information of test $t$ is necessary if we want to further reduce $T$; therefore, it is pointless to remove it.
Iterating this process over all tests of $T$ partitions the tests into two sets: a set of tests $N$ that are necessary, and a set of tests $P$ that can potentially be removed.
This process is described in \autoref{algo:find_req}.
Notice that it is quite costly because we need to do this for every test that can be potentially removed, but in practice, we hope that it reduces the number of elements we have to sample from. \new{It has complexity O($|\variants||N \cup P|^2|P|)$.}
\begin{algorithm} \small
    \caption{FindNecessary($P$, $N$, $\tests$): find the tests among a set of tests $P$ that are necessary in order to reduce $P \cup N$ for the test suite $\tests$}\label{algo:find_req}
    \begin{algorithmic}[1]
    \State \new{$P' \gets P$}
    \State \new{$N' \gets N$}
        \For{$t \in P$}
            \State \new{$C \gets (N' \cup P') \setminus \{ t\}$}
            \If{\textbf{not} $\solves_\tests(C)$}
                \State \new{$N' \gets N' \cup \{ t\}$}
                \State \new{$P' \gets P' \setminus \{ t\}$}
            \EndIf
        \EndFor
        \\ \Return \new{$N', P'$}
    \end{algorithmic}
\end{algorithm}

\paragraph*{Fast Elimination} The idea is to use a bisection algorithm as in \cite{bisection19}, for example, where it is used to find the relevant code locations.
Bisection algorithms are similar to binary search; the general idea is to use a divide-and-conquer approach where the search space is split, then tested, and based on the results of the test, it is reduced, and then the process repeats until one element is found. 
Here we are looking for one set among a set of test sets.
Here we have $P$, the set of tests from which we want to sample, $N$, the set of tests that are necessary, and $\tests$ the test suite considered, but the set of possible solutions is too large, making the bisection algorithm impractical.
Instead, the idea is to randomly divide $P$ into two sets $A$ and $B$ of the same size.
Then we can do the following: check if $A$ or $B$ is a solution with $\solves_\tests(A \cup N)$, if that is the case, we divide the size of $P$ by a factor of two, and we can repeat the process.
If none of them are solutions, then we make the following assumption: $A$ or $B$ is necessary; therefore, we compute the solution where $A$ is necessary and the solution where $B$ is necessary and return the best.
In each step, we divide the size of $P$ by two, which allows much faster progress.

\paragraph*{Algorithm}
This process is formalized in \autoref{algo:sample_sol}.
The first step is to use \autoref{algo:find_req} to reduce the variance, and then we can proceed with the sampling.
\new{Notice that in any recursive call to $BS$, the size of $P$ is divided by 2, therefore the longest call chain we can have is $O(\log |P|)$. In the worst case scenario, we make two recursive calls, therefore the worst case complexity is $O(2^{\log |P|}|\variants||N \cup P|^2|P|)=O(|\variants||N \cup P|^2|P|^2)$.
It is expected that it only has polynomial complexity since we do not try every possible combination, we use the bisection to make large cuts in order to provide solutions faster.}

\begin{algorithm} \small
    \caption{BS($P$, $N$, $\tests$): sample a solution to the RTSM problem for $\tests$}\label{algo:sample_sol}
    \begin{algorithmic}[1]
        \State \new{$P', N' \gets FindNecessary(P, N, \tests)$}
        \If{\new{$P'$ is empty}} \Comment{\new{Every test is necessary}}
            \State \textbf{return} \new{$N'$}
        \EndIf
        \State \new{$A, B \gets randomSplit(P')$} \Comment{\new{Split the non-necessary in two random sets of same size}}
        \If{\new{$\solves_\tests(A \cup N')$}}\Comment{\new{If it is a solution then we can keep reducing it}}
            \State \textbf{return} \new{$BS(A,  N', \tests)$}
        \ElsIf{\new{$\solves_\tests(B \cup N')$}}
            \State \textbf{return} \new{$BS(B, N', \tests)$}
        \Else \Comment{\new{No solution found, assumes that $A$ or $B$ is necessary and output the best}}
            \State \new{$S_A \gets BS(B, A \cup N', \tests)$}
            \State \new{$S_B \gets BS(A, B \cup N', \tests)$}
            \State \textbf{return} best of $S_A$ or $S_B$
        \EndIf
    \end{algorithmic}
\end{algorithm}

\subsection{Divide and Conquer}

\new{As mentioned previously, $\solves_{K \cup P}(K)$ has already quadratic complexity, and our $BS$ algorithm is in $O(|\variants||N \cup P|^2|P|^2)$ and since initially $P=\tests$ and $N= \emptyset$ then the worst case complexity is $O(|\variants||\tests|^4)$.} Therefore, a more scalable, practical approach must be used.

\begin{figure}[ht]
    \centering
    \vspace{-130pt}
    \begin{tikzpicture}[
tests/.style={circle, draw=green!60, fill=green!5, very thick, minimum size=5mm},
action/.style={rectangle, draw=red!60, fill=red!5, very thick, minimum size=5mm},
if/.style={rectangle, draw=blue!60, fill=blue!5, very thick, minimum size=5mm},
]
\node[action, align=center] (input)                                           {\textbf{Input}: \\set of tests $\tests$};
\node[action] (split)   [right=.25cm of input]                  {split};
\node[tests]  (tk)      [right=.25cm of split]                  {$T_k$};
\node         (tk_m1)   [above=.25cm of tk]                     {...};
\node         (tn_m1)   [below=.5cm of tk]                      {...};
\node[tests]  (t2)      [above=.25cm of tk_m1]                  {$T_2$};
\node[tests]  (t1)      [above=.25cm of t2]                  {$T_1$};
\node[tests]  (tn)      [below=.25cm of tn_m1]                  {$T_n$};
\node[action]  (solve_t1)      [right=.25cm of t1]                     {BS};
\node[action]  (solve_t2)      [right=.25cm of t2]                     {BS};
\node[action]  (solve_tk)      [right=.25cm of tk]                     {BS};
\node[action]  (solve_tn)      [right=.25cm of tn]                     {BS};
\node[tests]  (sol_t1)      [right=.25cm of solve_t1]                     {$S_1$};
\node[tests]  (sol_t2)      [right=.25cm of solve_t2]                     {$S_2$};
\node[tests]  (sol_tk)      [right=.25cm of solve_tk]                     {$S_k$};
\node[tests]  (sol_tn)      [right=.25cm of solve_tn]                     {$S_n$};
\node         (solk_m1)     [above=.25cm of sol_tk]                      {...};
\node         (soln_m1)     [below=.5cm of sol_tk]                      {...};

\node[action]  (merge1)      [right=.25cm of sol_t1,yshift=-.5cm]                {merge};
\node[action]  (merge2)      [right=.25cm of sol_tk,yshift=.4cm]                {merge};
\node[action]  (merge3)      [right=.25cm of sol_tn,yshift=.4cm]                {merge};

\node         (merge_dots1)  [right=.5cm of merge1,yshift=-1cm]                      {...};
\node         (merge_dots2)  [right=.5cm of merge2,yshift=-1cm]                      {...};
\node[tests]  (sol)  [right=1.25cm of merge2]                      {$S$};

\node[if] (if) [right=.25cm of sol] {\textbf{If} $S\neq \tests$};
\node[if] (else) [below=.25cm of if] {\textbf{Else}};
\node[action] (out) [below=.25cm of else] {\textbf{Output:} $S$};

\draw[->] (input.east) -- (split.west);
\draw[->] (split.east) -- (t1.west);
\draw[->] (split.east) -- (t2.west);
\draw[->] (split.east) -- (tk.west);
\draw[->] (split.east) -- (tn.west);
\draw[->] (t1.east) -- (solve_t1.west);
\draw[->] (t2.east) -- (solve_t2.west);
\draw[->] (tk.east) -- (solve_tk.west);
\draw[->] (tn.east) -- (solve_tn.west);
\draw[->] (solve_t1.east) -- (sol_t1.west);
\draw[->] (solve_t2.east) -- (sol_t2.west);
\draw[->] (solve_tk.east) -- (sol_tk.west);
\draw[->] (solve_tn.east) -- (sol_tn.west);
\draw[->] (sol_t1.east) -- (merge1.north west);
\draw[->] (sol_t2.east) -- (merge1.south west);
\draw[->] (sol_tk.east) -- (merge2.south west);
\draw[->] (solk_m1.east) -- (merge2.north west);
\draw[->] (sol_tn.east) -- (merge3.south west);
\draw[->] (soln_m1.east) -- (merge3.north west);
\draw[-] (merge1.east) -- (merge_dots1.west);
\draw[-] (merge2.east) -- (merge_dots1.west);
\draw[-] (merge3.east) -- (merge_dots2.west);
\draw[->] (merge_dots1.east) -- (sol.north west);
\draw[->] (merge_dots2.east) -- (sol.south west);
\draw[->] (sol.east) -- (if.west);
\draw[->] (if.north)  .. controls +(right:5mm) and +(up:85mm) .. (input.north) node [midway,xshift=-2.5cm] {restart with $\tests= S$};
\draw[-] (if.south) -- (else.north);
\draw[->] (else.south) -- (out.north);

    \end{tikzpicture}
    \caption{High-level overview of our divide and conquer with an iterative strategy used in our \prefixours algorithm}
    \label{fig:biss}
    \Description{The figure illustrates a classic divide and conquer approach on random subsets of the input test set except that if the solution subset is smaller than the original the algorithm is restarted with the solution.}
\end{figure}

The idea is to use a divide-and-conquer approach and to use sampling, here BS, as a subroutine. 
\autoref{fig:biss} illustrates at a high level our divide-and-conquer approach with our iterative strategy.
In the first step, we split $\tests$ into multiple smaller sets and \new{sample solutions for} these smaller instances of the RTSM problem.
From then on, at each iteration, we merge two solutions of the previous step.
\new{By merging, we mean that the RTSM instances are merged, but instead of starting from scratch we start as a base solution from the union of the two solutions found at the previous step.}
We keep iterating until there is only one instance left.
\new{The goal is that, when facing larger instances, the set $P$ of instances to potentially remove is much smaller reducing the complexity.}
In the worst case, in all the sub-instances, no better solutions than including every test of the sub-instance could be found; therefore, the last step is similar to not doing any divide and conquer.
The initial split thus has a large variance and, therefore, can have a large impact on the performance of this approach.

The process is described in detail in \autoref{algo:fusion_sol}. But before diving into the explanations of the algorithm, let us convince ourselves that this process gives a valid solution.
Given two solutions $A$ and $B$ for the RTSM instances $T_A$ and $T_B$, is $S$ a solution to the RTSM instance $A \cup B$ also a solution to the entire RTSM instance $T_A \cup T_B$?
The answer is no, the reason being that when we use $solve$, we perform a linear regression that depends on the exact true data.
\new{When we consider only $A \cup B$, we may be able to remove the tests, because we have the exact data to get the correct ranks, but the predicted performance cannot be fed as true data for another linear regression; it may be too inaccurate.}
To do this correctly at minimal cost, when a solution $S$ is found to the RTSM instance $A \cup B$ we check if it is also a solution to the entire RTSM instance $T_A \cup T_B$, if it is we accept it and continue otherwise we try again a certain number of times until we arbitrarily set $S$ to $A \cup B$.


\begin{algorithm} \small
    \caption{DC($\tests$, $n$, \new{$BS$}): from a set of tests $\tests$ by splitting it into $n$ parts and using \new{$BS$ a sampling strategy by default ours}, find a solution by using divide and conquer}\label{algo:fusion_sol}
    \begin{algorithmic}[1]
        \State $Q \gets split(\tests, n)$
        \State \new{$Q \gets \{ BS(q, \emptyset, q) \ | \ q \in Q \}$}
        \Comment{produce initial solutions for each sub-instance}
        \While{$Q$ is not empty}
            \State $(A, T_A), \ (B, T_B) \gets $ first two elements of $Q$
            \State \new{$S \gets BS(A \cup B, \emptyset, A \cup B)$}
            \If{\textbf{not} $\solves_{T_A \cup T_B}(S)$} \Comment{check if $S$ is NOT a valid merge}
                \State $S \gets A \cup B$ \Comment{discard the intermediate solutions}
            \EndIf
            \State $Q.append((S, T_A \cup T_B))$
        \EndWhile
        \State \textbf{return} $pop(Q)$
    \end{algorithmic}
\end{algorithm}

Finally, we combine our divide-and-conquer with \textit{ iterative solving}, since our method is based on random sampling.
Iterative solving solves the RTSM problem from an already existing solution and follows the idea that if a smaller cost test set is found, then we may try one more time to find another improvement until no improvement is found.

\section{Other Approaches}\label{sec:baselines}

We describe a few methods inspired by related work and intuition that we consider as baselines.
\new{Notice that \autoref{algo:fusion_sol} takes $BS$ as input, it implies that if we have sampling strategies, they can also be used in combination with this approach.}

Our method is called \prefixours. All methods, unless specified, use divide-and-conquer and iterative solving.

\paragraph{Random Search \prefix{random}}\label{sec:rs}
Random search is a simple baseline that can be easily adapted to the RTSM problem.
The random search involves randomly sampling a set of tests and then checking if the set forms a solution that can be returned at any time. 
Rather than sampling among all possible candidates, the process can be adapted to be iterative.
Initially, there are $n$ tests in the suite, then we randomly search for a set of tests of size $n-1$, \new{if a valid solution is found then we sample a set of tests of size $n-2$ from the solution of size $n-1$ and so on.}
\new{The complexity is $O(k|\variants|n^2$) where $k$ is the number of samples taken in the worst case.}

\paragraph{Greedy Algorithm \prefix{greedy}}\label{sec:greedy}
Another simple black-box baseline in a greedy search approach. 
At each step, we can try to remove the test that provides the least information; since all tests have the same weight, we chose to remove the test with the lowest metric variance because, on average, it should contribute the least to the ranking.
Note that we also tried with the lowest mean metric, and the results were approximately the same, so we decided not to report this variant for brevity reasons. \new{More formally, we remove a $t^* \in \text{argmin}_t \text{Var}(P_{*t})$ where $P_{*t}$ is the vector containing the performance of all the variants on $t$. We break ties by alphabetical order.}
\new{The complexity is $O(|\variants||\tests|^2$) in the worst case.}

\paragraph{Mixed Integer Linear Programming MILP}\label{sec:milp}

\new{We can transform the WRTSM problem into an approximate MILP formulation.}
Let us use the notations introduced in \autoref{def:wrtsm}: $\mathcal{A}$ is the set of variants, $\mathcal{T}$ is the set of tests, $p_{k,i}$ \new{is} the performance of variant $k$ on test $i$, $cost(t_i)$ is the cost of test $t_i$ and the weights of the solution $ {\color{red} w_i} \in \mathbb{R}^+$ for the test $t_i$.
Let us explain our encoding; the presence or absence of the test $t_i \in \mathcal{T}$ in the solution is encoded as ${\color{palblue} u_i} \in \{0, 1\}$.
Our constraints \new{to keep the same ranking} are encoded as follows:
\begin{align*}
    \sum_{i} {\color{palblue} u_i} \times p_{k,i} &\leq \sum_{i} {\color{palblue} u_i} \times p_{l,i} &\text{if variant $k$ is ranked lower than variant $l$}\\
    \sum_{i} {\color{palblue} u_i} \times p_{k,i} &\geq \sum_{i} {\color{palblue} u_i} \times p_{l,i} &\text{if variant $k$ is ranked higher than variant $l$}
\end{align*}
Thus, our objective is to minimize:
\begin{equation*}
   \min \sum_i {\color{palblue} u_i} \times cost(t_i)
\end{equation*}
However, this formulation does not take into account the weights {\color{red}$W \in \mathbb{R}^{|\tests| \times 1}$} of the tests in the solution. They can be included in the following way in our constraints for variant $k$ and variant $l$:
\begin{align*}
    &\sum_{i} {\color{palblue} u_i}  {\color{red} w_i}\times p_{k,i}   \leq \sum_{i} {\color{palblue} u_i}  {\color{red} w_i}\times p_{l,i} &\text{if variant $k$ is ranked lower than variant $l$}\\
    &\sum_{i} {\color{palblue} u_i}  {\color{red} w_i}\times p_{k,i} \geq \sum_{i} {\color{palblue} u_i}  {\color{red} w_i}\times p_{l,i} &\text{if variant $k$ is ranked higher than variant $l$}
\end{align*}
Notice that this is no longer a linear program, since both $ {\color{red} w_i}$ and ${\color{palblue} u_i}$ are decision variables.
Therefore, we use the previous linear version, without  the weights {\color{red}$W \in \mathbb{R}^{|\tests| \times 1}$}.
\new{This formulation solves a simpler version of the problem; therefore, the solutions of the MILP approaches are not optimal for the problem where the ${\color{red} w_i}$ can be set, although they are optimal for the ${\color{red} w_i}= 1$ cases.}
Observe that this formulation also only enables us to consider a Kendall of exactly 1, if we wanted to consider a Kendall of $T$, then we would use the following formulation for variant $k$ and variant $l$:
\begin{align*}
    &{\color{palgreen} c_{k,l}} \sum_{i} {\color{palblue} u_i} \times p_{k,i}   \leq \sum_{i} {\color{palblue} u_i} \times p_{l,i} &\text{if variant $k$ is ranked lower than variant $l$}\\
    &{\color{palgreen} c_{k,l}} \sum_{i} {\color{palblue} u_i}  \times p_{k,i} \geq \sum_{i} {\color{palblue} u_i} \times p_{l,i} &\text{if variant $k$ is ranked} 
\end{align*} with ${\color{palgreen} c_{k,l}} \in \{-1, 1\}$, they are also decision variables and the additional constraint that:
\begin{equation*}
    \frac{1}{2|\mathcal{A}|^2}\sum_{k,l \in \mathcal{A}} {\color{palgreen} c_{k,l}} \geq T
\end{equation*} since ${\color{palgreen} c_{k,l}} = 1$ for a concordant pair and ${\color{palgreen} c_{k,l}} = -1$ for a discordant pair, translating exactly the left-hand side into the definition of the Kendall.
This means that changing the target Kendall also implies that this becomes a non linear program, thus MILP is a very rigid approach \new{and we must use the limited original formulation}.
Note that MILP is used as a standalone baseline without divide-and-conquer.
In our implementation, MILP uses the COIN (CBC) solver~\cite{John2024coin}.

\paragraph{Principal Component Analysis \prefix{PCA}}\label{sec:pca}
Principal Component Analysis (PCA) is a technique used to reduce dimensionality while minimizing loss in terms of variance.
We use PCA on the performance matrix and look at the most important vector, \new{$v\in\mathbb{R}^\tests$}, the one who explains most of the variance.
Then we choose to remove one test that is the least important for this vector: \new{$t^* \in \text{argmin}_t v_t$, in case of ties, we take the smallest test by alphabetical order.}

\subsection{Failed Attempts}

In this section, we report other attempts to obtain some baselines that were not included in our experimental results.
We decided not to include them due to poor results compared to existing baselines but believe it is of scientific interest to report what we failed to make work.

\paragraph{Racing Algorithm}\label{sec:racing}

Based on racing algorithms such as \textsc{irace}~\cite{LOPEZIBANEZ201643}, we adapt their approach based on Friedman's test~\cite{Friedman01121937} and produce the same decisions \textsc{irace} would take.
In other words, it works iteratively and chooses a test to remove at each step.
The test chosen is based on Friedman's test used in \textsc{irace} to detect the poorest performers; here, our intuition is that it should detect the test that has the least discriminative power and remove it.
We implemented it in such a way that if we run \textsc{irace} on our setup, except that it does not have new data at each step, then \textsc{irace} would remove the same tests in the same order.
Applying it in practice over all our benchmarks led to no cost reduction at all on all seeds; therefore, this baseline is discarded.

\paragraph{Genetic Algorithm}\label{sec:ga}

There is a large body of work on the minimization of the test suite that uses genetic algorithms~\cite{harman2001search, harman2012RT}.
However, most of these approaches are white- or gray-box approaches. Unlike these approaches, there are neither features nor associated code; therefore, we cannot adapt their approaches.
We can still adapt the approach of genetic algorithms to our case.
Our genes consist of zeros and ones; a one indicates that we keep a test in the test suite, and a zero indicates that it is discarded. Then, we can frame the RTSM problem as a multi-objective optimization problem by considering two independent objectives, namely 1) minimizing the size of the test suite and 2) minimizing the ranking error so that it stays below the desired threshold.
Genetic algorithms, unlike other methods, cannot guarantee that the ranking error stays below the desired threshold; that is why we have to consider it as an objective function to minimize it.
In order for genetic algorithms to work with our two objectives, we use the following formula:
\begin{equation*}
    score(T) = \begin{cases}
        accuracy(T) \ \text{if $T$ is not accurate} \\
        10 + (1 - \frac{size(T)}{N})
    \end{cases}
\end{equation*}
When we say `$T$ is not accurate', we mean that it does not predict the classification of the variants to the required accuracy, that is, there is too much classification error.
The idea is that we favor the sets $T$ that match the required accuracy threshold and then rank them according to their size, since $\frac{size(T)}{N}$ describes the ratio of tests kept from the problem. It is close to a linear combination, but since we want a solution in a limited time, we observed that we need to bias for correct solutions with the desired accuracy, to always get solutions within the timeouts.

We tried changing the score to a linear combination or using a multi-objective genetic algorithm, and we tried different variations of the parameters of the genetic algorithm to no avail: we only found worse-performing versions.
Since we did not make genetic algorithms work correctly, they are not included in the experiments. 



    \section{Experiments}\label{sec:experiments}

We ask the following questions:
\begin{itemize}
    \item RQ1: How many tests can be eliminated while maintaining the same ranking?
    \item RQ2: How many tests can be eliminated while maintaining approximately the same ranking?
    \item RQ3: How is minimization affected with only a subset of the variants?
    \item RQ4: \new{How many benchmark runs are needed to break even between minimization cost and savings?} 
    \item RQ5: How does our method \prefixours compare to the others?
    \item RQ6: How do the different components affect performance?
\end{itemize}

\subsection{Experimental details}

We implemented our methods in Python; an archive of the code is available as supplementary material along with all additional figures and tables.
We will report Kendall's $\tau$ as mentioned previously (see \autoref{sec:definition}).
\new{There is no ground-truth data for the minimal benchmark, and it is prohibitively expensive to calculate. 
Our empirical evaluation focuses on the absolute reduction in the benchmarks and the relative ranking of the different algorithms considered.}
All experiments were run on a single core on Intel Xeon Gold 6130 (16 cores, 2017), and our algorithms returned the best result found so far. 
Working with multiple cores can speed up the results almost linearly, and better results may be achieved with a longer timeout on a single core, since all methods except MILP are anytime algorithms.
\new{Since our methods are stochastic, we conducted our experiments on 10 different seeds and performed statistical tests when necessary.}
The timeouts were 50 minutes per task for all datasets.
As a reminder, all methods use divide-and-conquer and iterative solving except MILP.
\new{In our experiments, there was no tie in the ranks of the variants.}

\new{One challenge acquired is that while theoretically, our divide and conquer should guarantee that the solution has the desired Kendall value, that is not the case. 
When two solutions $s_1$, $s_2$ for the tests suites $\tests_1$ and $\tests_2$ are merged, theoretically if they give perfect ranking respectively, then using $s=s_1 \cup s_2$ for $\tests=\tests_1 \cup \tests_2$ should give perfect ranking but because of the objective framed in the linear regression this does not happen and a worse result is obtained.}

\subsection{Datasets Used}

In general, we evaluate our methods on more than 50 benchmarks across a wide range of domains and software.

\paragraph{Code Generation: HumanEval~\cite{chen2021codex,evalplus}} First, we looked at the HumanEval benchmark~\cite{chen2021codex} and also considered the HumanEval Plus~\cite{evalplus} version with additional tests to correctly test generated code; furthermore, in the repository, the authors provide the samples generated by numerous models. We took these models as variants and the HumanEval tasks as tests, which gives us a matrix of zeros and ones depending on the pass $k$ score used. We took pass1 for both the classic version and the plus versions, giving us 2 benchmarks. These datasets contain 120 variants and 164 instances.
The cost for these benchmarks is 1 for all tests, as no cost was provided in the original data.

\paragraph{Algorithm Selection: ASlib~\cite{bischl_aslib_2016}} Second, we looked at comparisons of algorithm performance. There are a large number of scenarios in the ASlib benchmark~\cite{bischl_aslib_2016}, and we selected only scenarios where there were at least 8 variants.
In fact, when too few variants are available, it is not as relevant.
When multiple versions of the same data are available, we only keep the one with the most data.
This produced 23 benchmarks with one metric to optimize, and the mean running time of the test was used as a cost; this means that these are instances of the WRTSM problem.
The idea of using cost is to use the fact that some more complicated instances can provide more information but at the cost of more runtime than, for example, 10 smaller tests; therefore, not adding weights would be an unfair comparison of tests.
This leads to a range of diverse benchmarks, from answer set programming, constraint solving, SAT, and others, with fewer than 100 variants and fewer than 500 tests.
The number of variants is usually in the range of 8-20.

\paragraph{Configurable Systems~\cite{LesoilSGATBJ24,alvespereira:hal-02148791,kaltenecker2020interplay,lesoil2023}} Third, we looked at datasets that measure the performance of configurable systems~\cite{LesoilSGATBJ24,alvespereira:hal-02148791,kaltenecker2020interplay,lesoil2023}.
Specifically, the dataset contains 8 unique benchmarks, but multiple metrics are measured; therefore, we generated variants of the benchmarks with an increasing number of metrics, but we limited ourselves to 4 variants of the same benchmark.
This adds a pool of 19 additional benchmarks with up to 200 variants and thousands of tests.
For sqlite we made only 4 versions despite 15 different output metrics.
The cost in these benchmarks is the mean running time of the tests across all variants when available; if time is a metric, then it is used as cost. For example, gcc-ctime has ctime as a cost and metric, whereas lingeling-conflicts has cost 1 for all tests since no time is in the metrics.

\paragraph{Program Repair: RepairBench~\cite{repairbench}}Finally, we looked at the performance of LLM in automatic program repair with RepairBench~\cite{repairbench}, which contained two benchmarks at the time: gitbugjava (90 tests, 23 variants) and defects4j (484 tests, 22 variants). A variant here is an LLM.
Then, three versions of each benchmark were created based on the release dates of the LLMs. The first cut-off date is Oct. 2024 (9 variants left), and the second one is Jan. 2025 (17 variants left).
Adding a final six additional benchmarks.
The cost for these benchmarks is 1 for all tests, since the cost was global and not on a test basis.

\subsection{RQ1: How many tests can be eliminated while maintaining the same ranking?}

\begin{figure}[th]
  \begin{subfigure}[b]{0.5\linewidth}
    \centering
    \includegraphics[width=.99\linewidth]{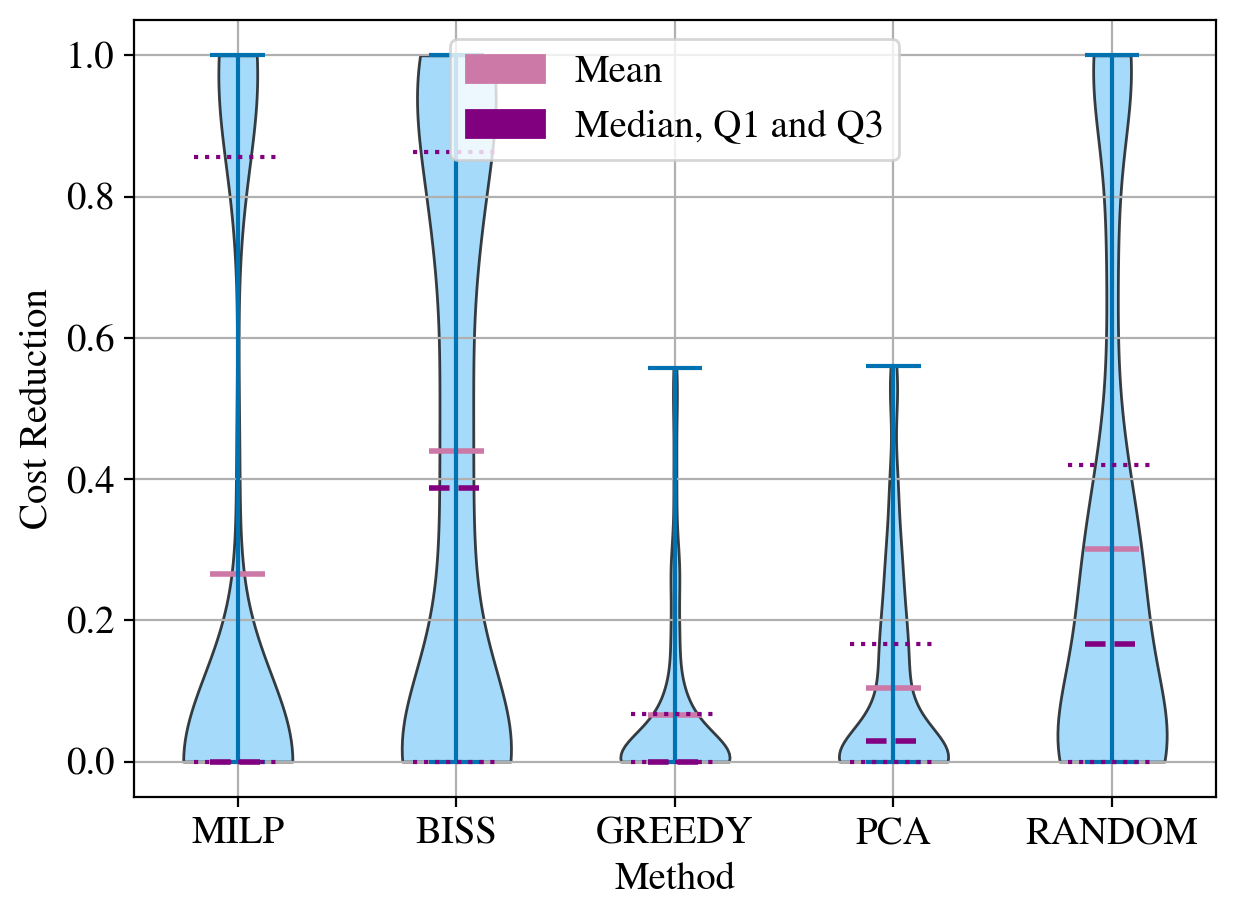} 
    \caption{All benchmarks} 
  \end{subfigure}
  \begin{subfigure}[b]{0.5\linewidth}
    \centering
    \includegraphics[width=.99\linewidth]{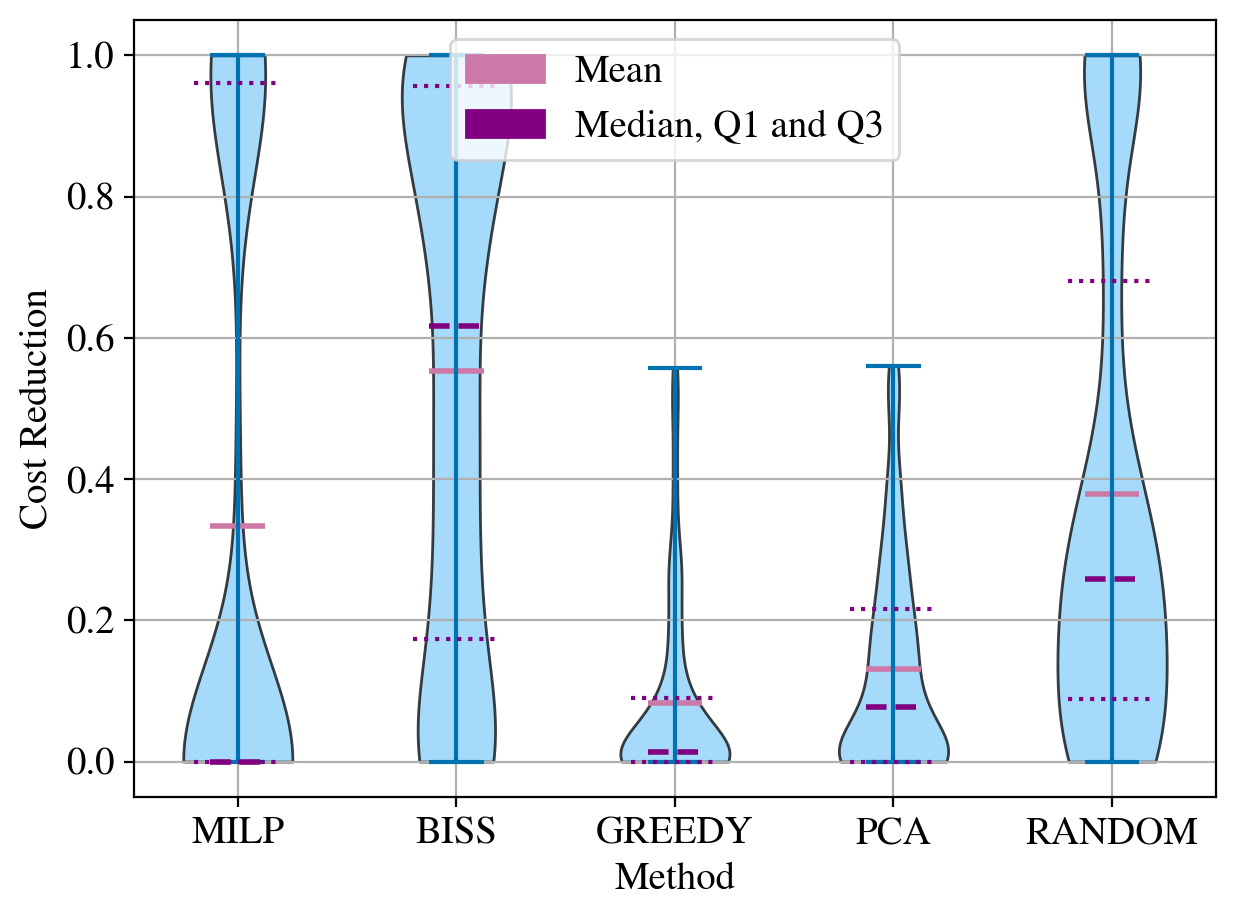} 
    \caption{Benchmarks with a cost reduction 39/50} 
  \end{subfigure} 
    \caption{Violin plot of the cost reduction (higher is better) of the different methods across different benchmarks with a target Kendall value of 1}\label{fig:gain}
\end{figure}

In order to tackle this question, we will run our methods on all datasets targeting no ranking error. 
\new{The aggregated results are plotted in a violin plot in \autoref{fig:gain}, the full results are in \autoref{table:gain_full} in the Appendix. The cost reduction is the fraction of cost saved, therefore higher is better.}

\new{For 11 of the benchmarks, no cost reduction was found, which means that for 39/50 we found a cost reduction.
We observe that globally \prefixours outperforms the other methods.
Interestingly the third quartiles of MILP and \prefixours are similar, meaning that they perform as well on the same fraction of benchmarks, however the worst case scenario for \prefixours is much better. 
MILP has a median at 0 cost reduction whereas \prefixours has a median nearly at 0.4.
\prefix{random} offers better average and median case than MILP. 
However, it offers a worse third quartile; in other words, it often finds reductions, but they tend to not be as important as those found by MILP.
\prefix{random} is on all aspects trailing behind \prefixours.
\prefix{Greedy} and \prefix{pca} are clearly outperformed by the others.}

\new{
When looking only at benchmarks where at least one method found a cost reduction, we clearly observe improved performances for all methods.
However, these improvements do not change the relative performance of one method compared to another and \prefixours still outperforms the other methods with the same arguments.
}

\begin{figure}[th]
  \begin{subfigure}[b]{0.5\linewidth}
    \centering
    \includegraphics[width=.99\linewidth]{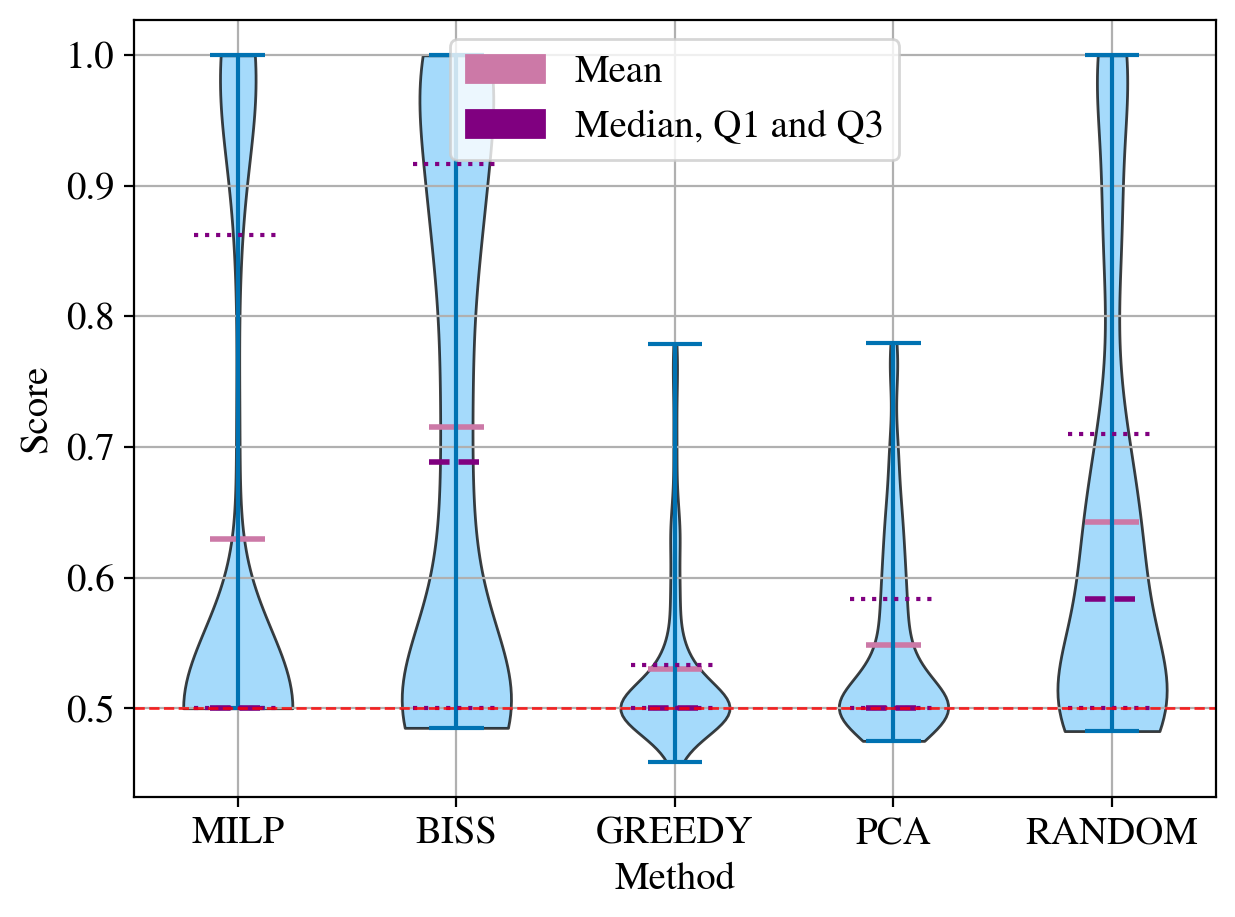} 
    \caption{All benchmarks} 
  \end{subfigure}
  \begin{subfigure}[b]{0.5\linewidth}
    \centering
    \includegraphics[width=.99\linewidth]{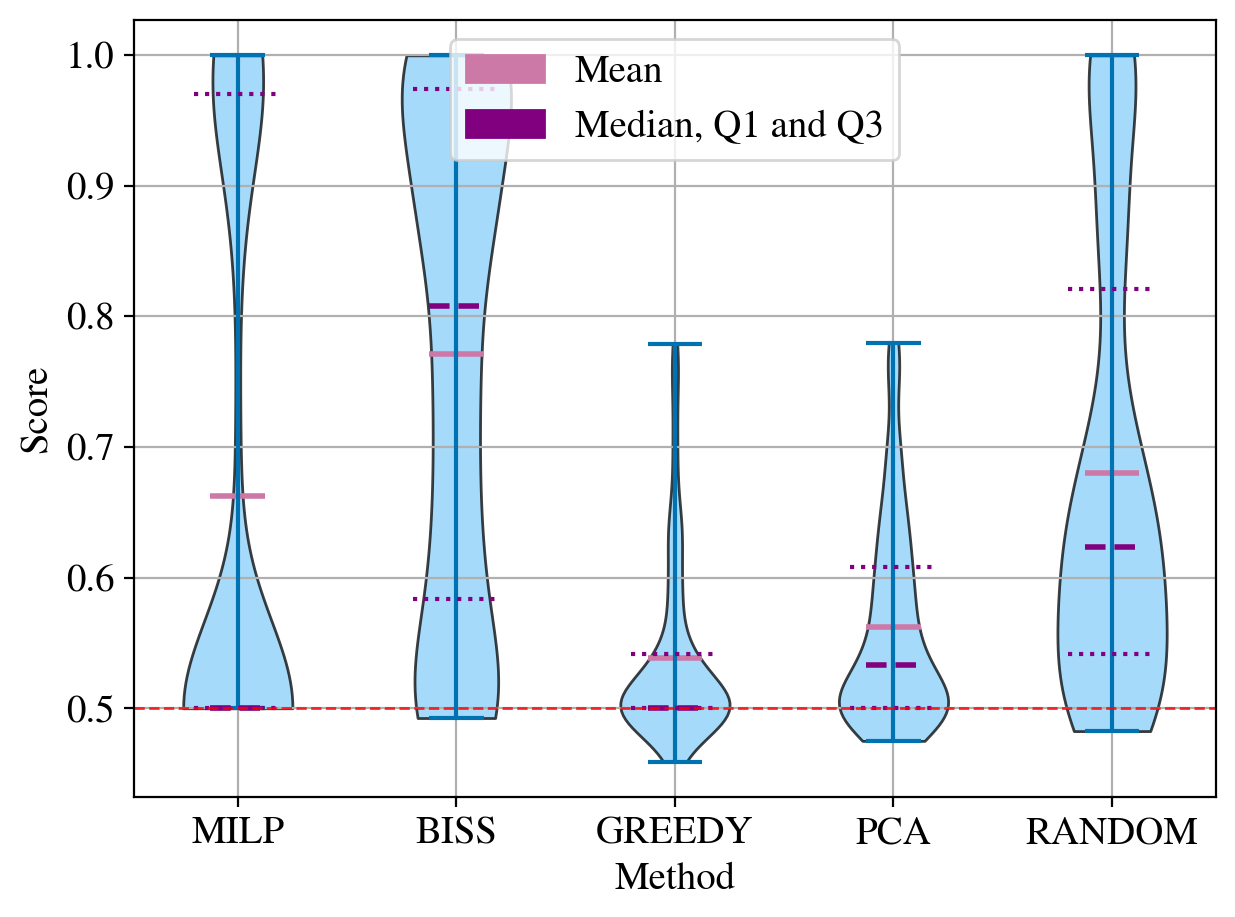} 
    \caption{Benchmarks with a cost reduction 39/50} 
  \end{subfigure} 
    \caption{Violin plot of the score (higher is better) of the different methods across different benchmarks with a target Kendall value of 1}\label{fig:gain_score}
\end{figure}

\new{
We found that in practice not all solutions have perfect Kendall.
This is due to our divide and conquer approach; without this component, the different methods have perfect ranking.
So we introduce a metric the score combining both the cost reduction and the Kendall value.
Intuitively, if the cost is reduced by 10\%, for an optimal trade-off, the Kendall value should not be more than 10\% lower.
The score gives equal weight to a cost reduction and a Kendall value reduction: \new{$score(c,k)= \frac{1}{2} (\frac{c}{2} + \frac{1 + k}{2})$} where $c \in [0;1]$ is the fraction of cost reduced and $k \in [-1; 1]$ is the Kendall coefficient.
By default, if no reduction is made, the obtained score is 0.5; another way to think about this is that if the score is above 0.5, then the trade-off starts to become interesting.
Therefore we plot in \autoref{fig:gain_score} the same data for the score and observe similar results as for the cost reduction.
}

\new{MILP by definition is the only method that guarantees a score of at least 0.5, which correspond to no cost reduction at all with the solution being the full test suite.
\prefixours is still the method which offers the best results and rarely scores less than 0.5.
}

\begin{tcolorbox}[boxsep=-2pt]
\new{In more than 75\% of the benchmarks, test instances were successfully removed with high Kendall.
Reductions of costs range from a few percent up to almost 100\% of the cost, with the best method averaging 44\% cost reduction on all benchmarks and 55\% on benchmarks where a reduction was found by any of the methods.
\prefixours outperforms other methods. MILP and \prefix{random} then outperforms \prefix{greedy} and \prefix{PCA} by far.}
\end{tcolorbox}

\subsection{RQ2: How many tests can be eliminated while maintaining approximately the same ranking?}

\begin{figure}[th]
  \begin{subfigure}[b]{0.5\linewidth}
    \centering
    \includegraphics[width=.99\linewidth]{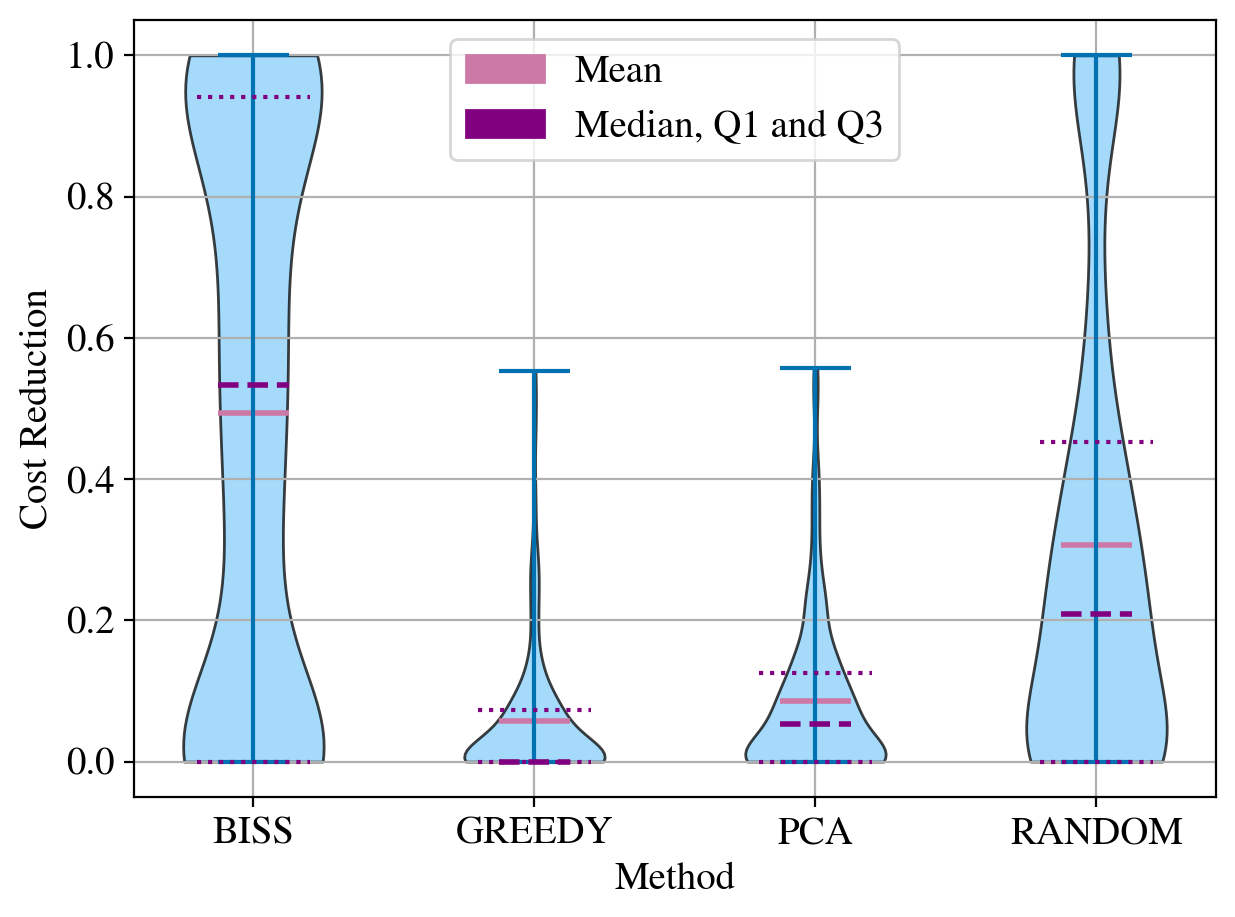} 
    \caption{All benchmarks} 
  \end{subfigure}
  \begin{subfigure}[b]{0.5\linewidth}
    \centering
    \includegraphics[width=.99\linewidth]{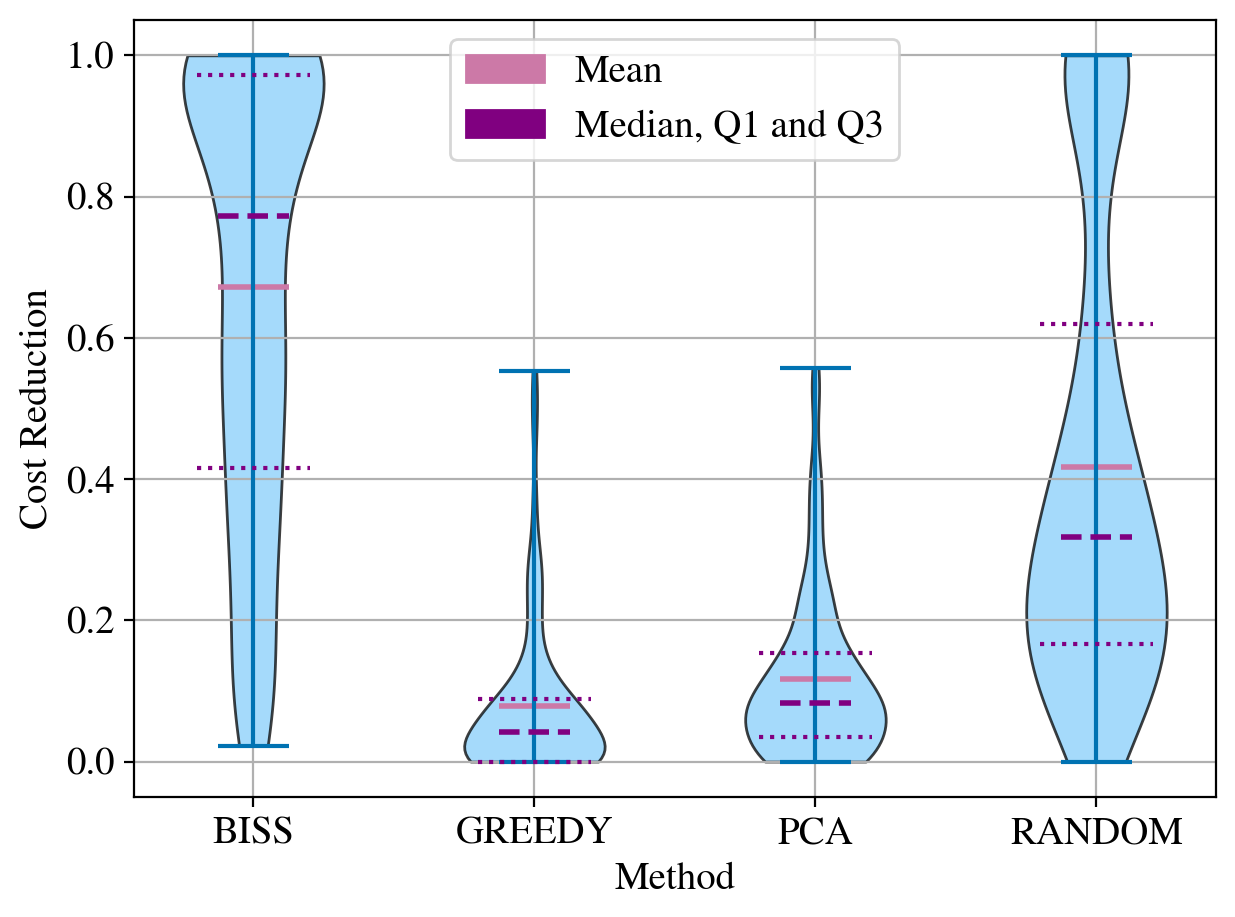} 
    \caption{Benchmarks with a cost reduction 36/50} 
  \end{subfigure} 
    \caption{Violin plot of the cost reduction (higher is better) of the different methods across different benchmarks with a target Kendall of 0.99}\label{fig:gain_99}
\end{figure}

For this RQ, we follow the same procedure as for RQ1, but with a Kendall target of 0.99, which allows for some leeway into the ranking while still being close to the original ranking. MILP is not present, because, as explained \new{before (\autoref{sec:milp}), we could not adapt the MILP formulation to target a Kendall value different from 1 within the MILP framework.}

\new{The aggregated results are found in \autoref{fig:gain_99} and the per benchmark results are available in the Appendix in \autoref{table:gain_99_full}.
First, no cost reduction was found on 14 benchmarks, a worse result than for RQ1. 
This is explained by MILP being absent which was the only method to find a cost reduction on some benchmarks.}

\begin{figure}[th]
  \begin{subfigure}[b]{0.5\linewidth}
    \centering
    \includegraphics[width=.99\linewidth]{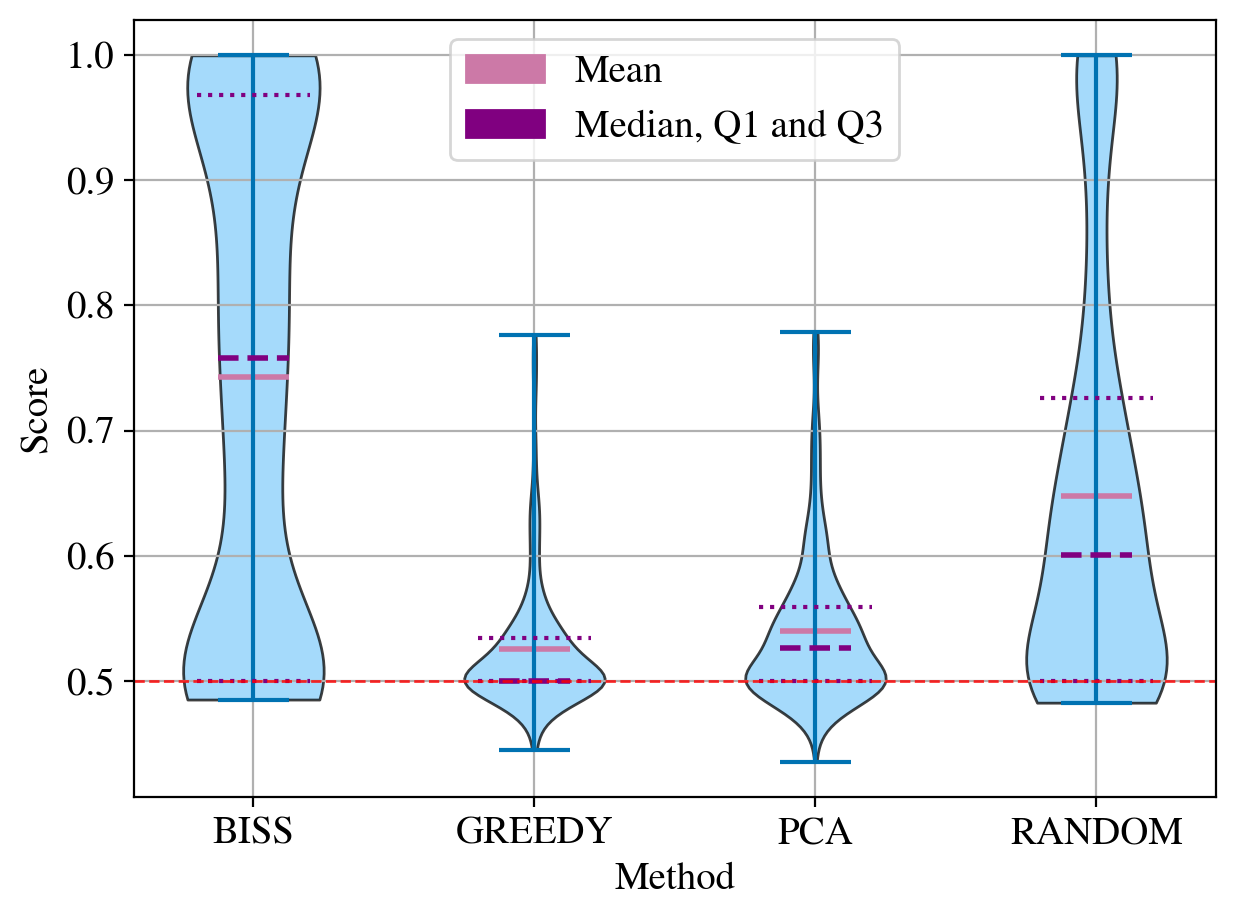} 
    \caption{All benchmarks} 
  \end{subfigure}
  \begin{subfigure}[b]{0.5\linewidth}
    \centering
    \includegraphics[width=.99\linewidth]{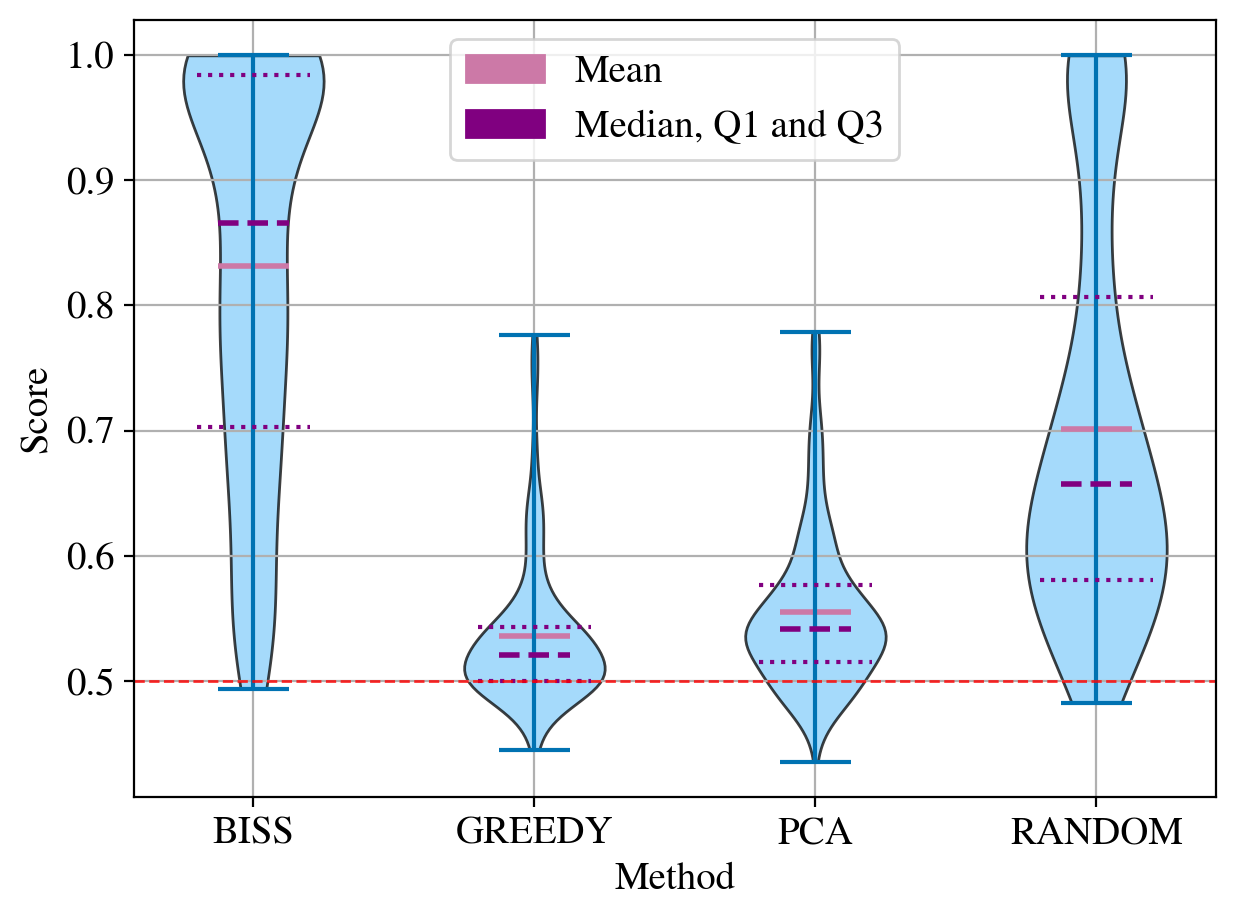} 
    \caption{Benchmarks with a cost reduction 39/50} 
  \end{subfigure} 
    \caption{Violin plot of the score (higher is better) of the different methods across different benchmarks with a target Kendall value of 0.99}\label{fig:gain_99_score}
\end{figure}

We observe the same trends as for \autoref{fig:gain}.
\new{\prefixours outperforms the others, followed by \prefix{random} and \prefix{greedy} and \prefix{PCA} have similar performances.
The mean cost reduction for \prefixours on all benchmarks is at 50\% and nearly 70\% on benchmarks where a cost reduction was found.
This clearly highlights that reducing a bit the constraints by targeting a lower Kendall value can offer significant improvements in performances.
}
\new{As in RQ1, we als plot the same data for the score in \autoref{fig:gain_99_score}.}

\new{And we observe the same conclusions as in \autoref{fig:gain_score}. \prefixours still outperforms all other methods.
All other methods are hurt by the reduced target Kendall value and have more solutions with scores < .5 than in RQ1 whereas it is not the case for \prefixours which has less of them.
}

\begin{tcolorbox}[boxsep=-2pt]
\new{The same relative performances trends are observed as in RQ1.
\prefixours still outperform the others.
\prefixours averages 50\% cost reduction on all benchmarks and almost 70\% on benchmarks where a reduction was found by any of the methods.
Reducing a bit the target Kendall value enables leeway offering much larger reductions.
Less benchmarks where found with a cost reduction despite the improvements because MILP was the only one to find improvements on these benchmarks.
}
\end{tcolorbox}

\subsection{RQ3: How is minimization affected with only a subset of the variants?}

\begin{figure}[ht] 
  \begin{subfigure}[b]{0.5\linewidth}
    \centering
    \includegraphics[width=.99\linewidth]{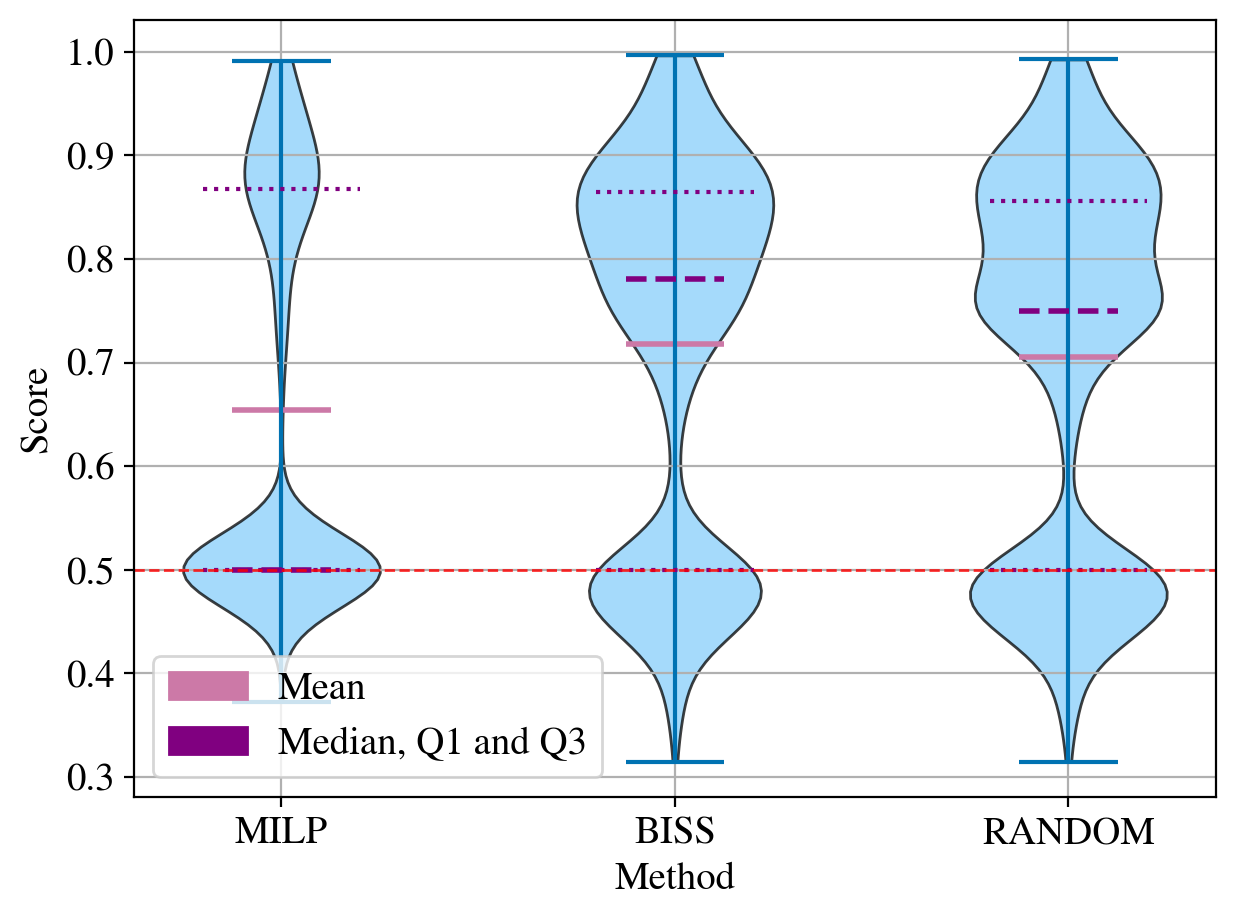} 
    \caption{25\% of variants} 
  \end{subfigure}
  \begin{subfigure}[b]{0.5\linewidth}
    \centering
    \includegraphics[width=.99\linewidth]{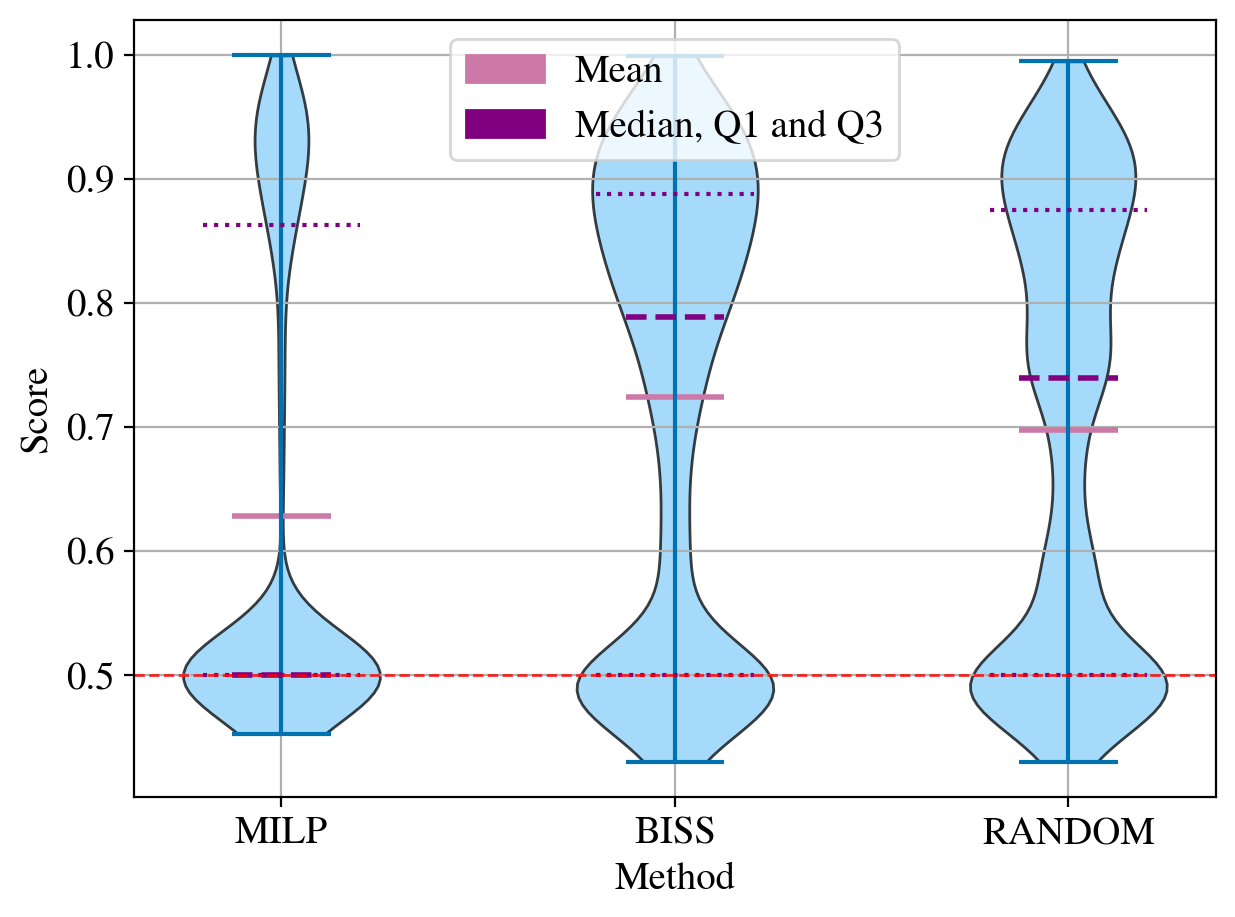} 
    \caption{50\% of variants} 
  \end{subfigure} 
  \begin{subfigure}[b]{0.5\linewidth}
    \centering
    \includegraphics[width=.99\linewidth]{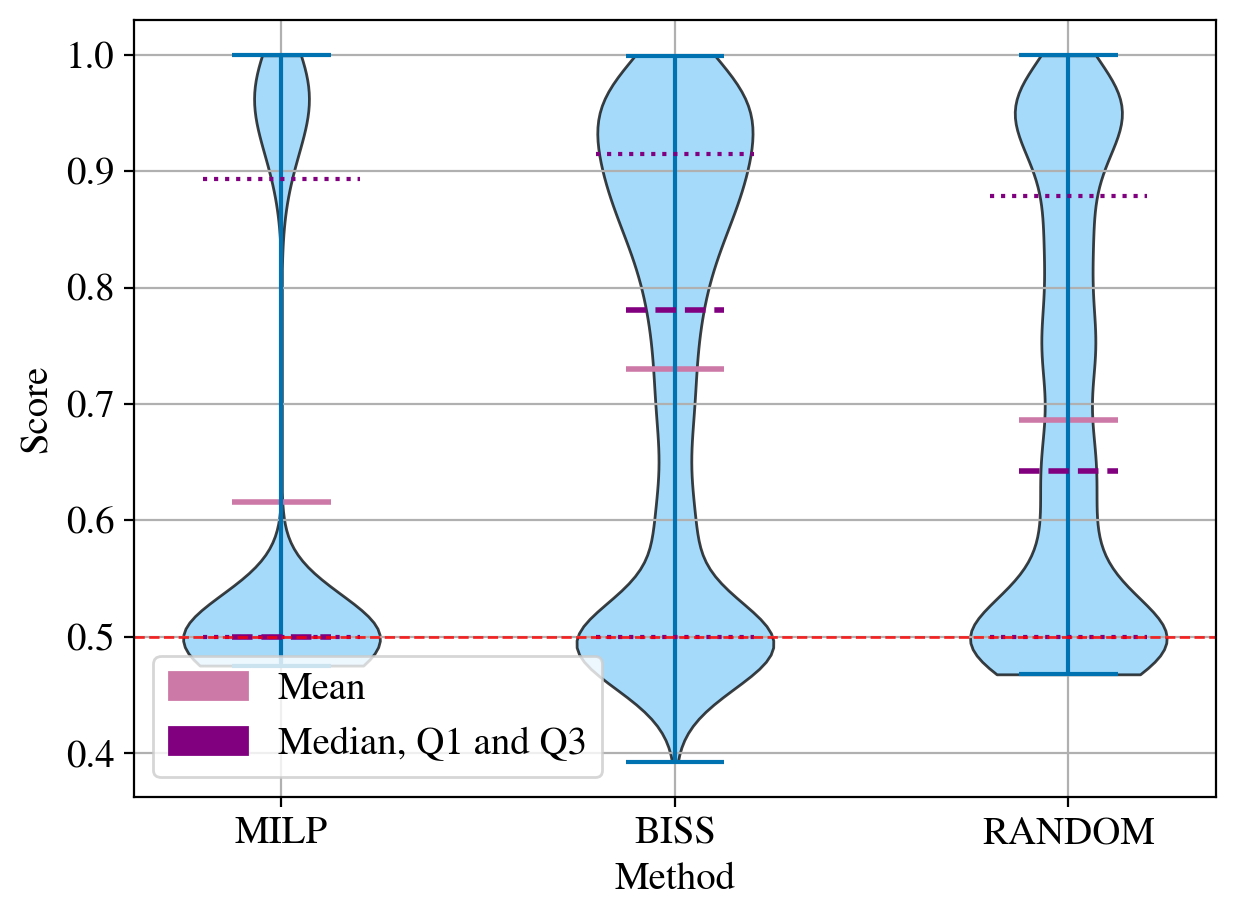} 
    \caption{75\% of variants} 
  \end{subfigure}
  \begin{subfigure}[b]{0.5\linewidth}
    \centering
    \includegraphics[width=.99\linewidth]{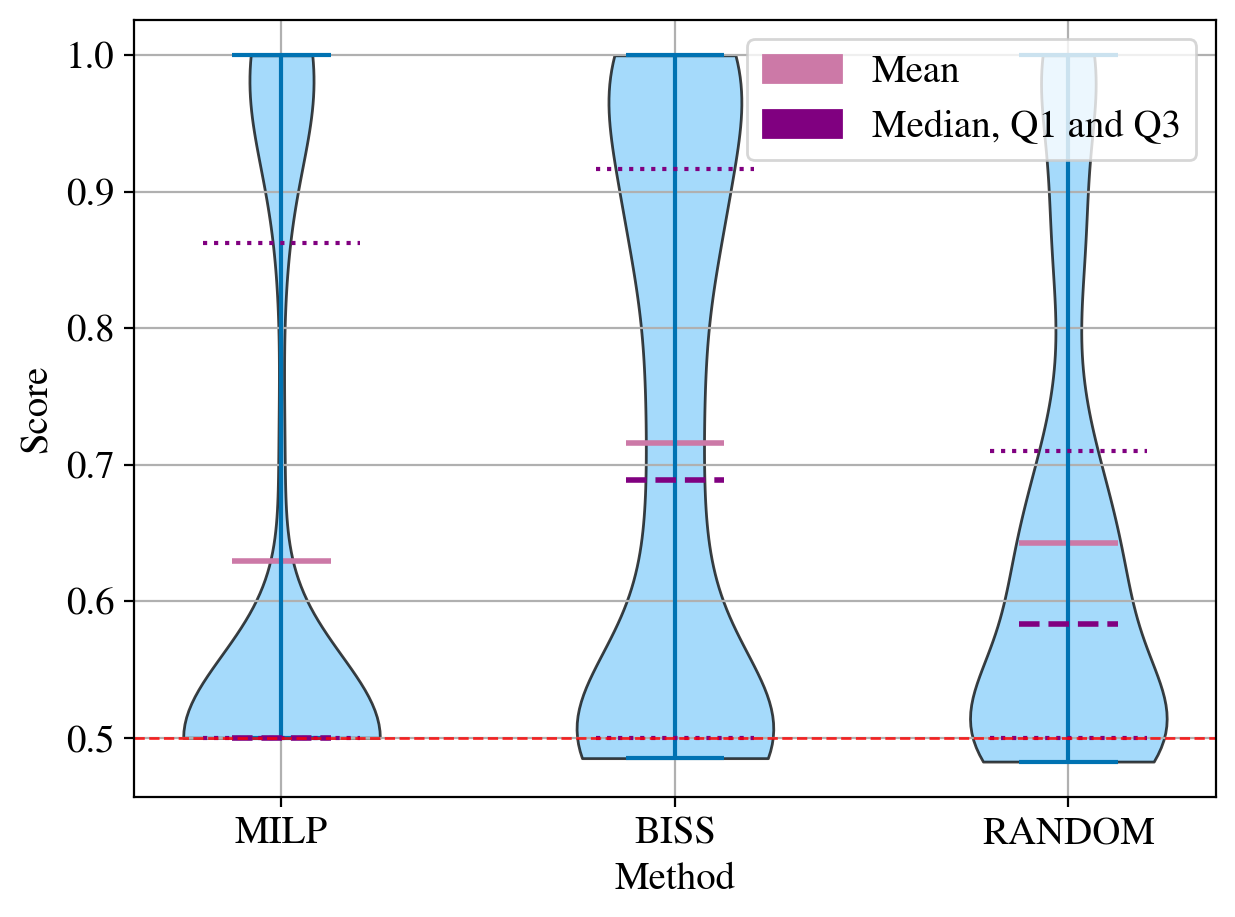} 
    \caption{100\% of variants} 
  \end{subfigure} 
  \caption{Violin plot of the score (higher is better) of the different methods across all benchmarks with a target Kendall of 1 for different fraction of variants. The score is measured against the data with 100\% of the variants. A score < 0.5 is worse than not doing any minimization, this is indicated by the \textcolor{red}{red} dotted line which draws the limit.}\label{fig:cum_dist_score}
\end{figure}

For this RQ, we only consider MILP, \prefixours and \prefix{random}, \new{the best performing methods for a target Kendall of 1}.
We sampled a fraction of all the variants for each benchmark, then ran the minimization procedure, and measured the cost reduction and Kendall on the full set of variants.
Unlike RQ2, we only consider targeting a Kendall of 1, the same ranking, because we are already losing information by removing variants.
This experiment reproduces \new{the context of} a competition using our approach. 
The benchmark is minimized with only a subset of all variants, then new variants are submitted by competitors, and the reduced benchmark is used.
We chose to keep 75\% 50\%, and 25\% of all variants. 
We sampled 10 random subsets for each percentage, \new{with 10 seeds per subset and method, that is, 100 seeds in total per method}. 

However, there are two metrics to consider: the cost reduction and the Kendall.
\new{We measure the Kendall on the benchmark with 100\% of the variants so the obtained Kendall may be much lower than the targeted one. }

Since there are 50 different benchmarks, \new{the supplementary material enables to plot individual data, here we will plot the aggregated data as violin plots of the score in \autoref{fig:cum_dist_score}.} 
\new{\prefixours generally outperforms MILP and \prefix{random}. 
}
\new{
Interestingly, when the number of variants is reduced the differences between \prefixours and \prefix{random} start to fade.
That means, when the number of variants is reduced, it is more likely that less tests are necessary to fully rank them. 
In that case it is then easier to find subsets of tests that provide benefits; in other words, it becomes easier to sample them.
When sampling good subsets of tests is easy, using \prefixours and \prefix{random} makes little difference, when it is harder the difference becomes important.
}

\new{Overall, MILP rarely provides solutions with a score < 0.5 compared to \prefixours and \prefix{random}; despite that its overall performance still trails behind \prefixours and \prefix{random}.
}

\begin{tcolorbox}[boxsep=-2pt]
\new{
 \prefixours still outperforms other methods despite the changes in variants.
 The less variants there are the more gains can easily be obtained.
 When few variants are used; it becomes easier to find solutions, differences between the performances of \prefixours and \prefix{random} fade.
 }
\end{tcolorbox}

\subsection{RQ4: \new{How many benchmark runs need to be made to make the minimization worth it?}}

\begin{figure}[th]
  \begin{subfigure}[b]{0.5\linewidth}
    \centering
    \includegraphics[width=\linewidth]{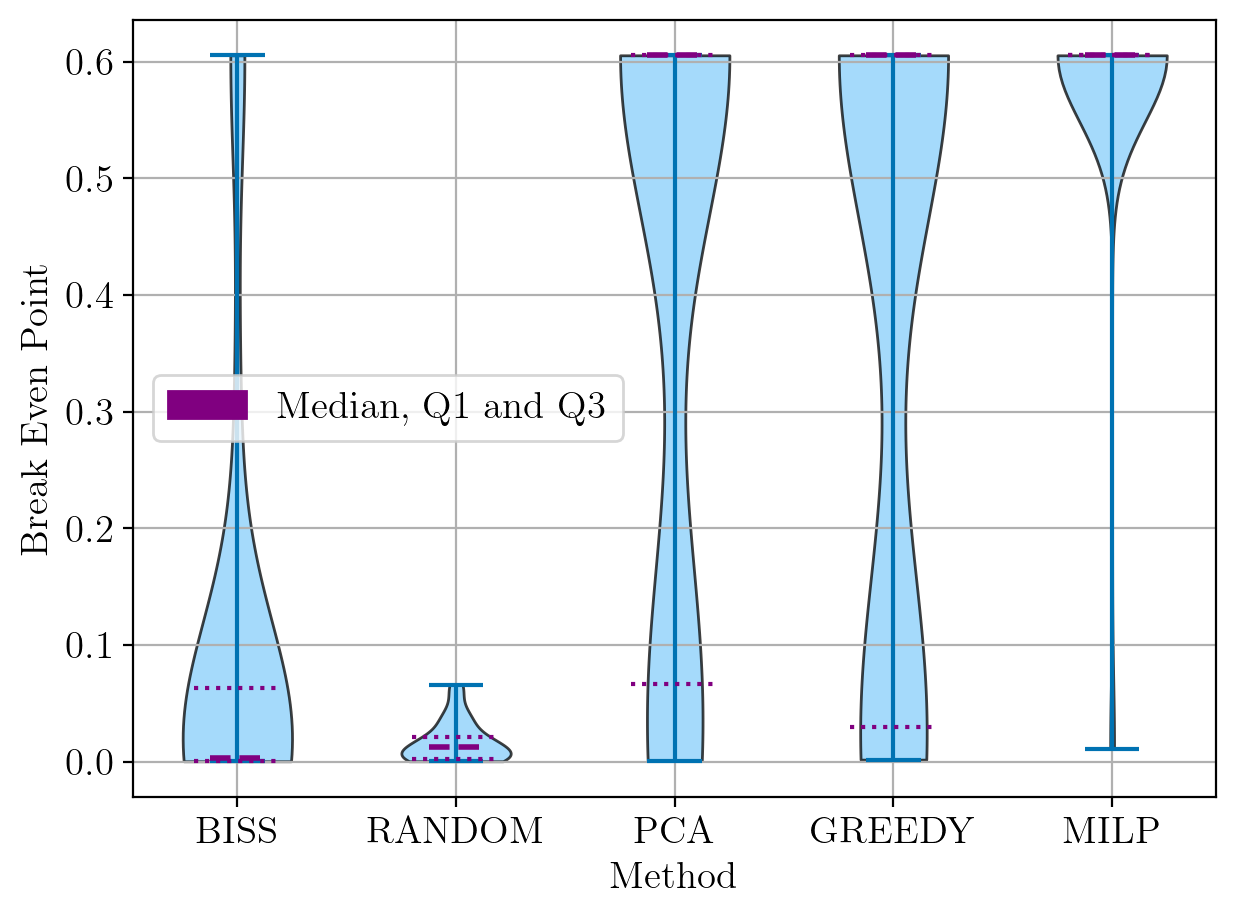}
    \caption{Target Kendall value : 1} 
  \end{subfigure}
  \begin{subfigure}[b]{0.5\linewidth}
    \centering
    \includegraphics[width=\linewidth]{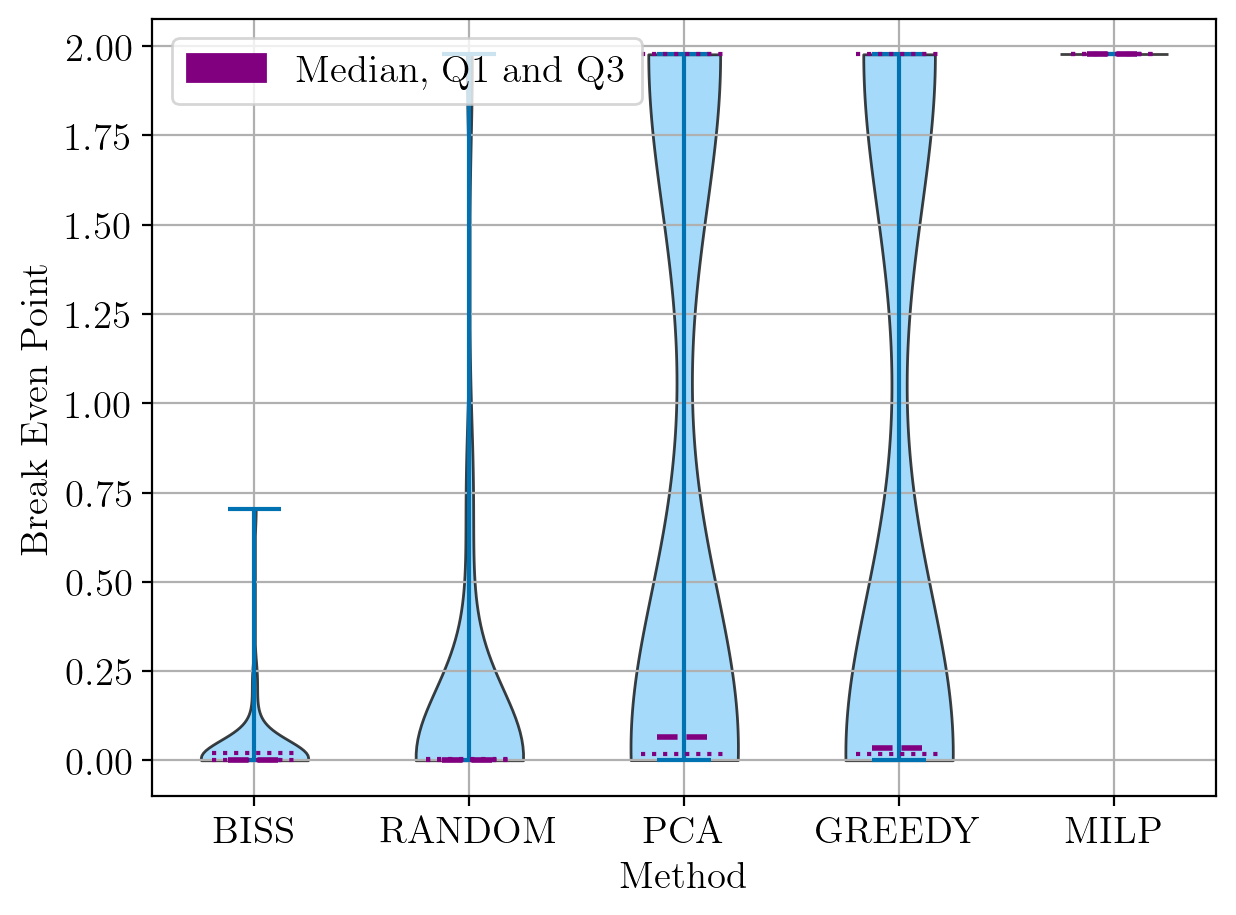}
    \caption{Target Kendallvalue 0.99} 
  \end{subfigure} 
    \caption{Violin plot of the break even points (lower is better) across the 12 benchmarks with costs for different target Kendall values with 100\% of the variants. Break even points translate how many times the minimized benchmark needs to be run to offset the cost of the minimization. All values here are less than 1, indicating that after one run, cost savings have already been made.} \label{fig:bep}
\end{figure}
\new{To answer this RQ, we use the same data as in RQ1 and RQ2, we will compute the break even point.
The break even point is the number of runs $n$ such that the cost of running the minimization is offset by the $n$ runs of the benchmark, the lower the better. More formally:}
\begin{equation*}
    \new{BEP(\tests', \tests, M) = \frac{\text{cost of minimization using $M$}}{\mathbb{E}[cost(\tests)] - \mathbb{E}[cost(\tests')]}}
\end{equation*}
\new{where $\tests'$ is the subset of tests returned by $M$ for the minimization of the test suite $\tests$.
This metric has no sense for benchmarks where no cost was defined therefore we exclude these, leaving us with 12 benchmarks.
When no cost reduction was found, we replace the data point with the worst break even point among all methods among all 12 benchmarks, this avoids severely unbalanced data points and penalize methods for not finding any improvements.
We plot the distributions on \autoref{fig:bep} for target Kendall of 1 and 0.99..
Technically, since we cannot run a benchmark partially, we can take the ceil of the break even point to get an integer number of minimized benchmark run; however that would leave no information on the Figure.
Indeed, all our methods, even the baselines provide minimizations that are offset after less than 1 run of the minimized benchmark.
}

\new{MILP is heavily penalized by the fact that most of the times it found no cost reduction and therefore is the worst performing method.}
\new{
\prefixours and \prefix{random} are the two best methods across the different target Kendall.
\prefix{random} score points by being very fast compared to \prefixours despite finding worst cost reductions.
Note that this does not take into account the obtained Kendall.}

\begin{tcolorbox}[boxsep=-2pt]
 \new{For all methods the cost of minimization is offset by less than one and at most two runs of the benchmark, highlighting the impact and potential savings of the RTSM problem.
 \prefixours and \prefix{random} outperform all other methods.}
\end{tcolorbox}
\subsection{RQ5: How does our method \prefixours compare to the others?}

Using data from RQ1 and RQ3, we performed statistical tests and measured the effect size when comparing the scores of our method with those of all other methods.
We ran a one-sided Wilcoxon signed-rank test to compare the scores of the different methods and also report the rank biserial correlation coefficient, that is, the effect size:
\begin{itemize}
    \item \prefixours outperforms \prefix{random} with effect size of 0.52 and p-value < $10^{-277}$
    \item \prefixours outperforms MILP with effect size of 0.24 and p-value < $10^{-308}$
    \item \prefix{random} outperforms MILP with effect size of 0.22 and p-value < $10^{-173}$
\end{itemize}
\prefixours outperforms all other methods, \prefix{random} outperforms MILP, and then MILP outperforms the other methods.

\begin{tcolorbox}[boxsep=-2pt]
 \prefixours outperforms all other methods; however, we note on a few benchmarks that using other approaches may lead to better solutions; more precisely, the MILP approach offers new perspectives for some benchmarks, whereas \prefix{random} fails on the same benchmarks as \prefixours.
\end{tcolorbox}

\subsection{RQ6: How does the different components affect performance?}

\begin{figure}[ht] 
    \centering
    \includegraphics[width=.7\linewidth]{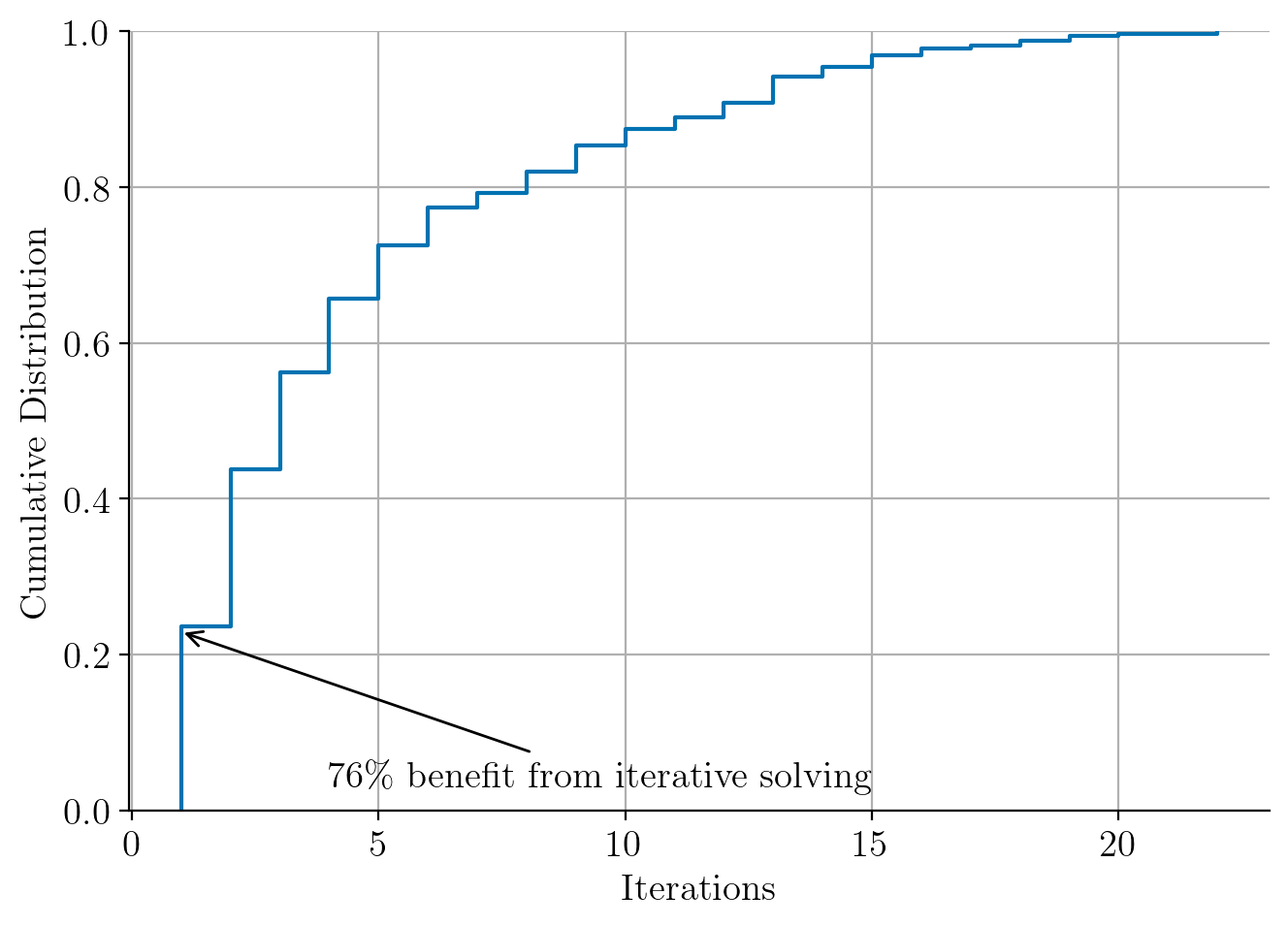} 

  \caption{Cumulative distribution of iterations for \prefixours with a target Kendall of 1 with 100\% of variants on the 33 benchmarks where a cost reduction was found. Smaller Area Under the Curve is better.}\label{fig:crunch_loops}
\end{figure}

\paragraph*{Iterative solving}
In order to measure the impact of iterative solving, we counted each time iterative solving improved the final solution over all our benchmarks for \prefixours with a Kendall target of 1 with 100\% of variants, so a subset of the data from RQ1.
We plot the cumulative distribution of the number of iterations performed on \autoref{fig:crunch_loops}.
We found that for most benchmarks iterative solving enables us to find better solutions in 76\% of the cases where a cost reduction is found.
Iterative solving is useless when the best solution is found during the first iteration, usually when a scenario is either very easy with only a few tests remaining or when a scenario is very hard with little to no tests being removed.

\paragraph*{Timeouts with Divide and Conquer}
As part of our experiments, we recorded current solutions even in case of timeouts; when a timeout occurs, we use the full set of tests as a solution. 

In order to measure the impact of divide and conquer on the speed of our method, for all methods with divide and conquer on all different experiments we recorded the total number of timeouts.
Then we looked for the number of timeouts of \prefixours and \prefix{random} without divide and conquer and iterative solving on the same scenario as RQ1, that is with all variants and with a Kendall of 1, this constitutes 1099 runs compared to the 58445 that we have with divide and conquer.
As seen in Table~\ref{table:timeouts}, there is a large variation in the number of timeouts.
Note that since MILP is not seed dependent, fewer runs are explaining the difference compared to the number from raw number to percentage.

First, not using divide and conquer multiplies by 2 the number of timeouts in number while having x50 time less runs, so not using divide and conquer multiplies by x100 the chance of timeout.
Second, the best performing methods, \prefixours and \prefix{random}, have few timeouts (<1\%), MILP timeouts rarely.
\prefix{PCA} surprisingly times out quite often.


\begin{table}[]
\caption{Timeouts per method used over all different seeds. No DC stands for the same results without divide and conquer and iterative solving. A "/" stands for no data recorded. Note that only one experiment was run with No DC hence the different percentages. Since MILP does not depend on a seed there are also less runs for MILP.}\label{table:timeouts}
\begin{tabular}{@{}lrrrrr@{}}
\toprule
Method & \multicolumn{1}{l}{\prefixours} & \multicolumn{1}{l}{\prefix{random}} & \multicolumn{1}{l}{\prefix{PCA}} & \multicolumn{1}{l}{\prefix{greedy}} & \multicolumn{1}{l}{MILP} \\ \midrule
Raw Number   & 144                    & 40                    & 5093                    & 0                       & 2650                      \\
Percentage     & 0.25\%                   & 0.07\%                   & 8.73\%                    & 0.00\%                       & 4.54\%                     \\ 
\midrule
Raw Number (No DC)   & 295                    & 288                    & /                    & /                       & /                      \\
Percentage (No DC)     & 26.84\%                   & 26.21\%                   & /                    & / & /                     \\ \bottomrule
\end{tabular}
\end{table}

\begin{tcolorbox}[boxsep=-2pt]
 Iterative solving improved the solutions in 76\% of the benchmarks where we found a cost reduction. 
    Divide and conquer reduces chances of timeouts by a factor of 100.
    Overall, for the best performing methods, \prefixours and \prefix{random}, there are very few timeouts with divide and conquer (<1\%) whereas there is one in four run that timeouts without it.
\end{tcolorbox}

\section{Selected Case Studies}\label{sec:case_studies}

In this section, we will look deeper into what we can learn from our solution:
    \begin{enumerate}
        \item BNSL: Is the dataset highly redundant?
        \item SAT 18 and SAT 20: How does the test feature distribution change after minimization?
        \item defects4j: Can benchmark minimization tell us about the evolution of LLMs' skill on program repair?
    \end{enumerate}

\subsection{Case Study: BNSL: Is the dataset highly redundant?}

\begin{figure}[hb]
    \centering
    \includegraphics[width=.7\linewidth]{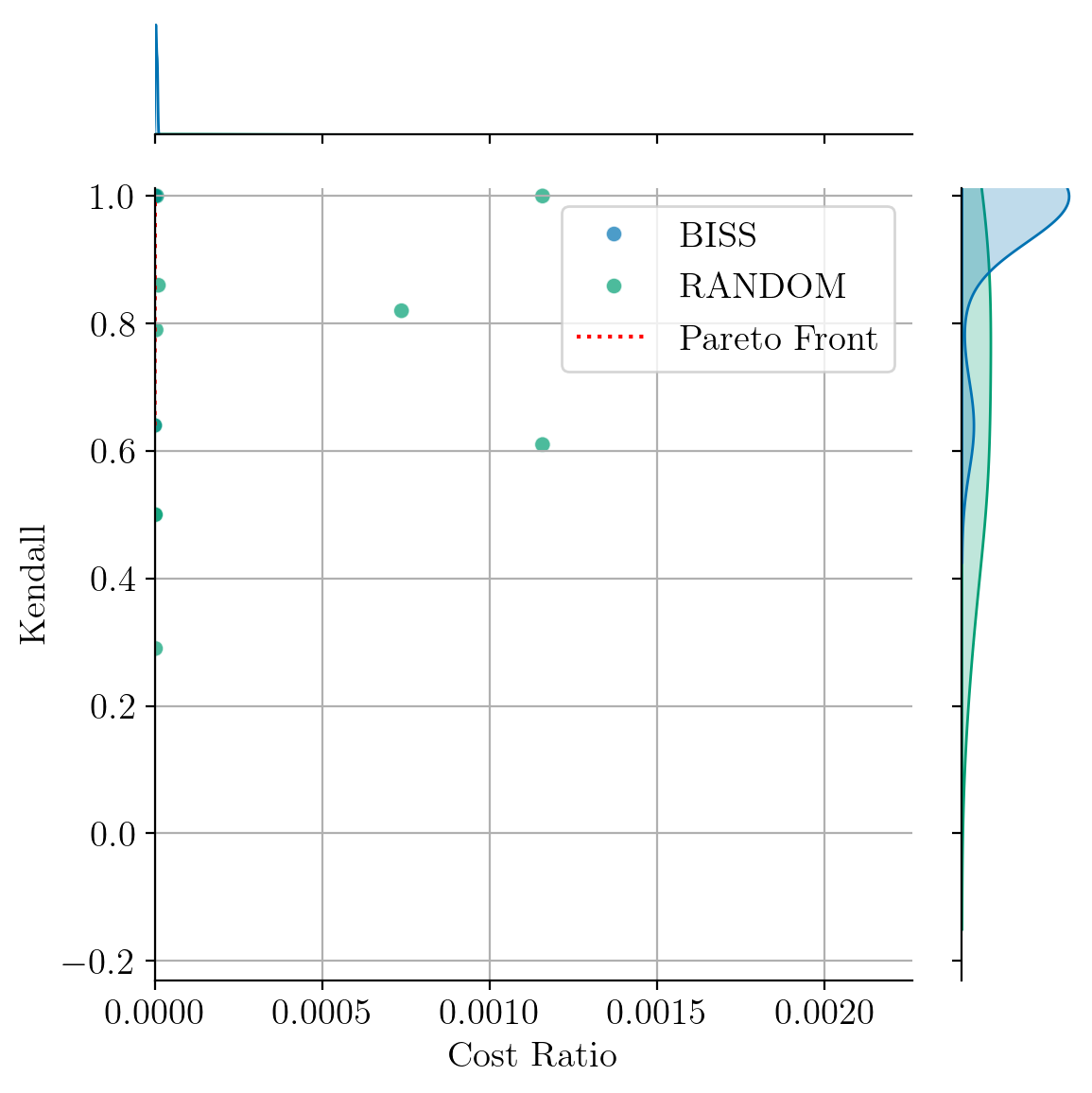} 
\caption{Kendall (higher is better) with respect to cost ratio (left is better) for different methods for BNSL-2016 with a target Kendall value of 1}\label{fig:bnsl-2016_cost}
\end{figure}

\new{BNSL is a benchmark for score-based structure learning in Bayesian networks, the performance measured is the runtime.
It comes from ASlib~\cite{bischl_aslib_2016} the algorithm selection collection of benchmarks.}
\new{We chose to target BNSL since} in \cite{matricon21comparison}, the authors find that it is easy to predict \new{the best performing solver among a challenger and an incumbent using only a single test} and emit the hypothesis that the benchmark is highly redundant.
We want to confirm this hypothesis; therefore, we will look more precisely at the BNSL data extracted from our methods \prefixours and \prefix{Random} the top performing ones on this scenario.

\new{
In \autoref{fig:bnsl-2016_cost} we plot the data points obtained for both methods and observe that all data points have very small cost ratios which basically amount to running a single test.
\prefixours' data points are agglomerated in the top left corner with a Kendall value fo 1 indicating that it is possible to have a perfect rankign with a single instance.
Even \prefix{random} finds good reductions even though with more variance.
}

Overall, looking at the data, their benchmark is highly redundant as was hypothesized in \cite{matricon21comparison}.


\begin{tcolorbox}[boxsep=-2pt]
 BNSL, as hypothesized in \cite{matricon21comparison}, is a highly redundant dataset with a ranking of variants that can be done with only one test instance. 
\end{tcolorbox}
\subsection{Case Study: SAT 18 and SAT 20: How does test features distribution change after minimization?}

In this case study, we want to observe how the different \new{distributional properties of the} features of the tests \new{change after minimization by \prefixours.}
We use the 10 different solutions with 100\% of variants obtained with \prefixours \new{from RQ1}.
\new{Intuitively, the hope is that the minimization does not impact much the distributional properties of the features of the test suite such as min, max, etc.}
We use the features given in \cite{bischl_aslib_2016} that were used for automated algorithm selection tasks.
\new{The protocol is the following: for each feature, we look at the relative change of distributional properties of the feature before and after minimization.
The distributional properties selected are: min, 1st quartile, median, 3rd quartile, max, mean, and std.}
\new{Since there are a lot of different features, we compute the median relative change across all features across all seeds.
Therefore for each distributional property we have a list of median relative change, one per seed, which we again aggregate by min, median, mean and max on the rows.
Then in \autoref{table:sat18_evol} and \autoref{table:sat20_evol}, we show this evolution for SAT18 and SAT20.
}

\new{For SAT18, a lot of test are easily removed, at least 38\%, but usually more than 75\%. 
Despite that, the relative change of the other distributional properties are all less than 11\%, except for Q3 and max. It is to be expected, since we want to minimize the cost, and instances with large features are often the instances, which take the most time therefore are more likely to be removed, therefore affecting max and Q3.
}
\new{For SAT20, some tests are easily removed, at least 24\% and usually more than 30\%, despite that, the relative change of the other distributional properties are all less than 9\%.
}
\new{On both benchmarks, the relative change in features is minor compared to the changes made in the test suite.}

\begin{table}[htb]
    \caption{Min, median, mean and max of relative median change across all features for solutions found by \prefixours on SAT 18}\label{table:sat18_evol}
    \centering
    \begin{tabular}{@{}ccccccccc@{}}
        \toprule  & \% of tests kept & mean & std & min & Q1 & median & Q3 & max\\
\midrule
min     & 38.81\% & 4.38\% & 4.26\% & 0.00\% & 3.25\% & 5.77\% & 7.03\% & 0.00\%\\
median  & 75.78\% & 8.71\% & 7.62\% & 0.00\% & 4.72\% & 7.60\% & 16.16\% & 6.95\%\\
mean    & 68.22\% & 8.39\% & 7.34\% & 0.82\% & 4.94\% & 7.71\% & 17.88\% & 9.01\%\\
max     & 84.99\% & 12.98\% & 10.59\% & 8.18\% & 8.50\% & 10.33\% & 36.87\% & 26.43\%\\
\bottomrule
    \end{tabular}

\end{table}

\begin{table}[htb]
    \caption{Min, median, mean and max of relative median change across all features for solutions found by \prefixours on SAT 20}\label{table:sat20_evol}
    \centering
    \begin{tabular}{@{}ccccccccc@{}}
        \toprule  & \% of tests kept & mean & std & min & Q1 & median & Q3 & max\\
\midrule
min     & 24.41\% & 3.62\% & 3.13\% & 0.00\% & 0.67\% & 1.17\% & 1.60\% & 0.00\%\\
median  & 30.74\% & 4.12\% & 4.40\% & 0.00\% & 1.51\% & 1.93\% & 3.04\% & 0.00\%\\
mean    & 37.51\% & 4.85\% & 5.05\% & 0.00\% & 1.87\% & 2.70\% & 3.91\% & 1.11\%\\
max     & 71.38\% & 8.60\% & 8.92\% & 0.00\% & 3.58\% & 7.98\% & 8.81\% & 6.97\%\\
\bottomrule
    \end{tabular}

\end{table}

\begin{tcolorbox}[boxsep=-2pt]
 \prefixours keep distributional properties of most features of the tests, with most relative changes largely smaller than the ratio of tests kept.
\end{tcolorbox}
\subsection{Case Study: Defects4j: Can benchmark minimization tell us about the evolution of LLMs' skill on program repair?}

\begin{figure}[hb]
\begin{subfigure}[b]{0.5\linewidth}
    \centering
    \includegraphics[width=\linewidth]{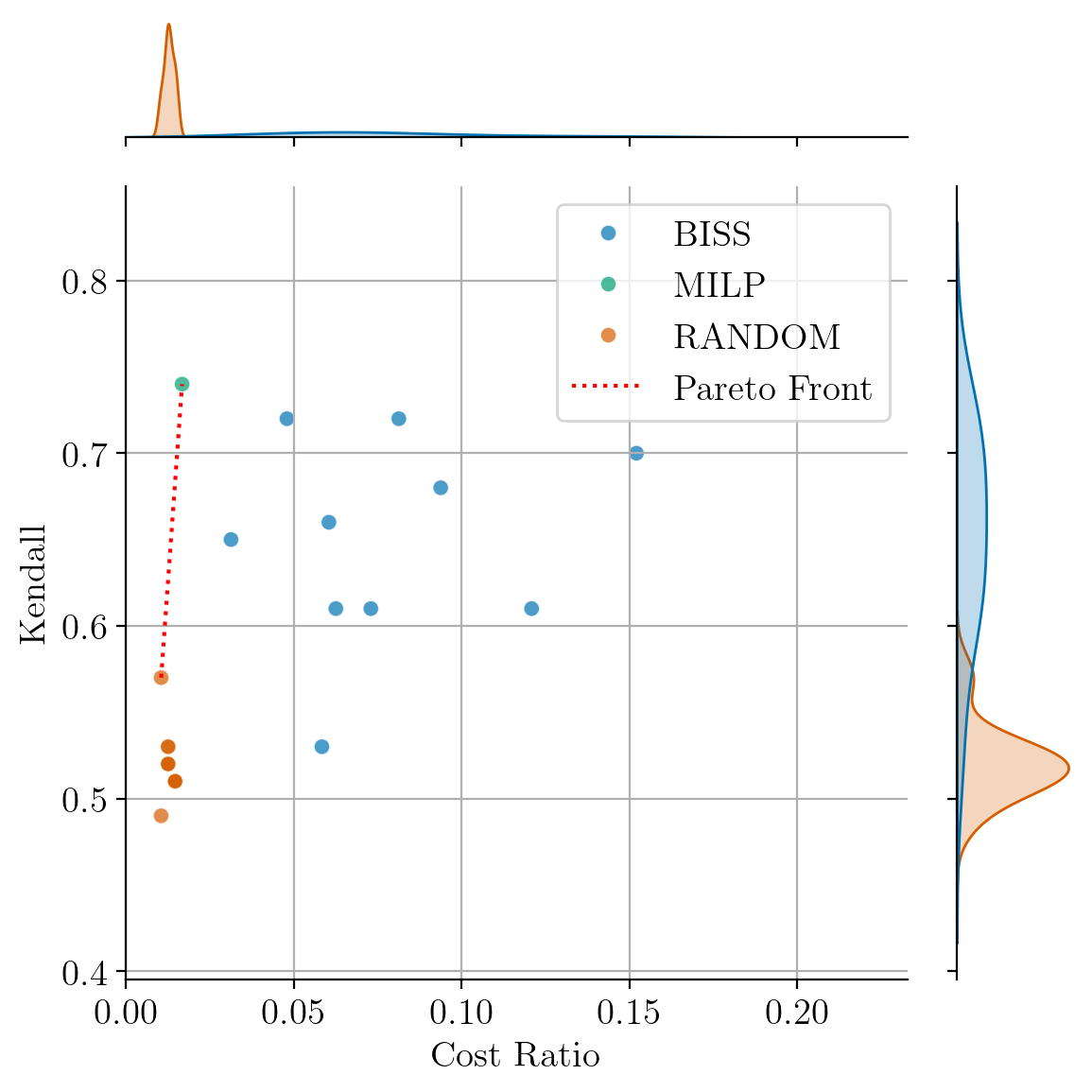} 
    \caption{cutoff on 2024/10/01 (9 variants/LLMs)}
  \end{subfigure}%
\begin{subfigure}[b]{0.5\linewidth}
    \centering
    \includegraphics[width=\linewidth]{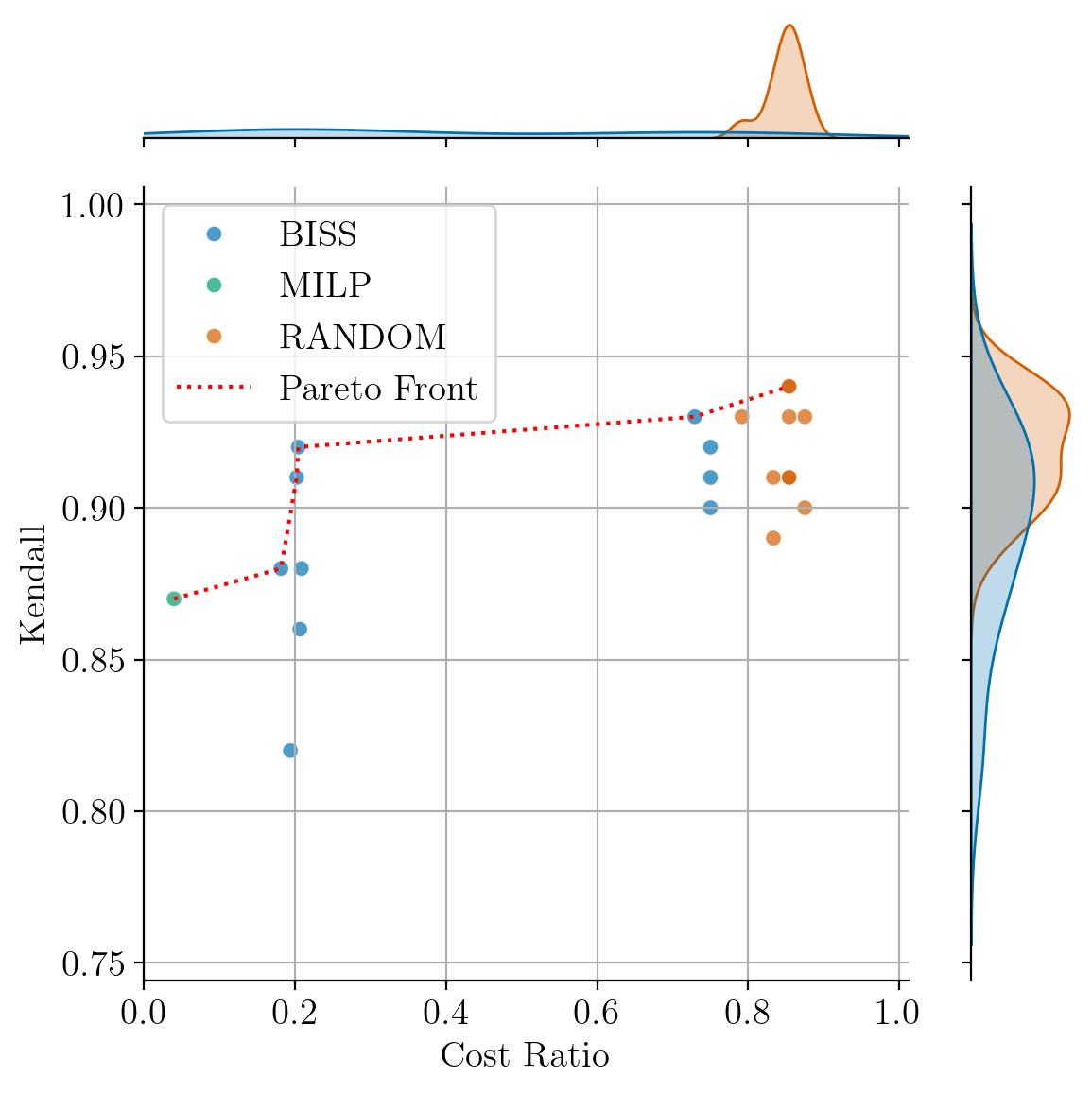} 
    \caption{cutoff on 2025/01/01 (17 variants/LLMs)}
  \end{subfigure}%
\\\begin{subfigure}[b]{\linewidth}
    \centering
    \includegraphics[width=.5\linewidth]{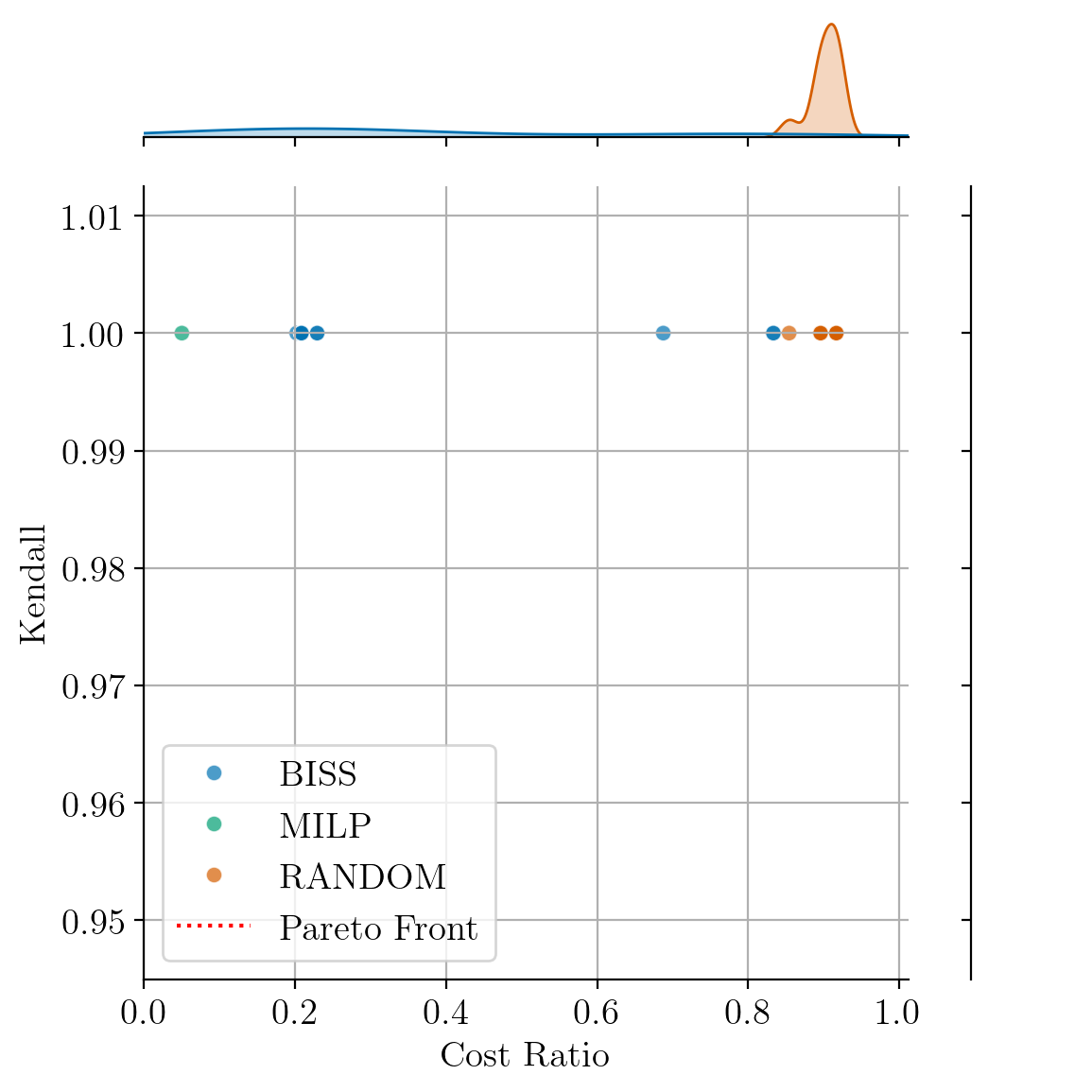} 
    \caption{no cutoff (22 variants/LLMs)}
  \end{subfigure}%
\caption{Defects4J Kendall (higher is better) with respect to cost ratio (lower is better) with 100\% of variants but with different cutoff dates  with a target Kendall of 1}\label{fig:defects4j}
\end{figure}

\new{RepairBench from Silva and Montperrus~\cite{repairbench} is based on Defects4J~\cite{defects4j} to measure the program repair capabilities of LLMs}.
To evaluate the evolution of the skills of LLMs, here, the variants, we plot on \autoref{fig:defects4j} the different data points of Kendall values with respect to the cost ratio for our methods.
The Kendall value is measured on the full dataset, that is, with all LLMs.
Here, the cost ratio is the ratio of tests kept in the minimized benchmark.

Interestingly, for the cutoff date in 2024, we observe a very low cost ratio, indicating that most LLMs before this date can be easily ranked.
The Kendall value, however, is quite low, indicating that the behavior of the more recent LLMs cannot be correctly captured by the present minimized test set.

\new{
For the cutoff date in 2025, we observe that the Kendall value jumps significantly, translating the fact there has been a lot of progress of LLMs on the task. 
Note that at that date, there were no `reasoning' LLMs; this is why we also observe that the Kendall \new{value is} not yet close to 1, translating that the reasoning behavior is not captured by the previous models.
In terms of cost ratio, we observe heterogeneous results that are actually quite similar to the cost ratio distribution of the case with no cutoffs. }

Finally, with no cutoffs, we observe that we can have cost reductions of more than 90\% of the tests set with no loss of accuracy in the best case, indicating that the benchmark is quite redundant for measuring the performances difference among the LLMs considered.
With MILP, we obtain a cost ratio of 5\%, meaning that all the LLMs' performances can be correctly ranked with only 5\% of the tests, which can save hundreds of dollars per attempt according to the costs reported by \cite{repairbench}.

\begin{tcolorbox}[boxsep=-2pt]
 Defects4J contains a lot of redundant instances that bring little to no discriminative power to rank the different LLMs.
 \new{Indeed,} 5\% of the 484 bugs are enough to perfectly rank all the LLMs, saving hundreds of dollars per attempt. \new{Interestingly, RepairBench has also been discontinued accord to their website\footnote{\url{https://repairbench.github.io/} last consulted Dec. 2 2025}.}
 The evolution of the cutoff dates and the results show that LLMs before Oct. 2024 behaved quite differently from the ones after, the LLMs released between Oct. 2024 and Jan. 2025 are more similar in terms of performance to the ones released afterwards.
\end{tcolorbox}

    \section{Threats to Validity}
\label{sec:threats}

\textbf{Internal threats.} 
\paragraph*{Timeouts} The configuration of timeouts presents a potential threat to the internal validity of our study, but since most methods are anytime, this leaves as much time as needed for the different methods.

\paragraph*{Stability}\label{para:stability} While it is theoretically true that adding more tests to a solution can only improve it, in practice, this is not always the case. We have observed that adding more tests can sometimes lead to instability. For example, our $FindNecessary$ procedure (see \autoref{algo:find_req}) may incorrectly identify certain elements as necessary when they are not. Unfortunately, we do not yet have a solution to this problem. However, our experimental results suggest that it is a minor issue. Our best experimental method is based on the $FindNecessary$ procedure, and the results show that the impact of this issue is very low.

\paragraph*{Generalizability} The effectiveness of our method relies on several hyperparameters, particularly within the divide-and-conquer part. These hyperparameters include the number of splits used to manage variance and the number of tests conducted during the iterative solving phase. While these parameters were selected manually, we avoided overfitting and extensive hyperparameter tuning to maintain the generalizability and robustness of our method across different datasets and scenarios. By selecting reasonable default values, we aimed to balance performance and applicability. 
Besides, we repeated each experiment ten times to account for variability and random effects. This repetition did introduce variance in the results. We reported and analyzed the standard deviations across different runs as part of our findings. 


\textbf{External threats.} The applicability of our study to real-world situations is influenced by the types of data we used for the evaluation. We tested our method using specific benchmarks from well-known sources, but this may not be representative of all possible datasets and domains. 
Our evaluation spans different domains (LLM, SAT, video, ...), with variants being independent systems or configurations from the same system, and different scenarios (leaderboard, competitions, performance modeling). 
An empirical question remains regarding the extent to which benchmarks can be minimized in each domain while still maintaining accurate rankings. Addressing this question is an ambitious task that would require extensive experiments. Our paper highlights the need for further exploration and assessment across broader datasets. We hope that other researchers will build on our work and conduct additional studies.



     \section{Discussion and Conclusion}\label{sec:impact}

In general, this analysis shows that some benchmarks are more amenable to minimization than others, highlighting the importance of domain specificities when using \prefixours. 

\textbf{Recommendations for benchmark design.} Benchmark designers, developers, creators can use our methods to design, understand, and refine their benchmarks. Our method automates this process by retaining the most relevant ones and, as shown, effectively removing tests that are too easy or repetitive. 
 It can be applied to existing benchmarks, effectively compressing them. We recommend using \prefixours since it generally outperforms other methods, and in case of bad results, to use MILP, which behaves quite differently. If benchmark designers want to test new variants on the reduced dataset (\eg to verify the hardness of the benchmarks), they should replicate the RQ2 experiments to ensure representativeness and control ranking error. 
As experiments show, on average better results are achieved when most of the variants remain in the problem. 

\textbf{Recommendations for software developers.} Developers consume benchmarks for performance comparison, nonregression, or tuning. Smaller benchmarks with the same discriminative power mean that consumers will maintain accuracy while spending less time executing their software on variants. For scenarios like tuning or nonregression, where benchmarks are run repeatedly, the gains are amplified with each run, resulting in substantial overall benefits. We thus offer the same recommendations as for benchmark design, and empirical results show that important gains can be achieved for reducing the computational cost. 

In addition, our method enables strategic targeting of the most performance-critical tests. 
 While general recommendations remain, developers who need to focus on a very few instances can even reduce the target Kendall to find even smaller representative instances at the cost of accuracy. They can then improve their systems before testing on the larger benchmark. However, there is a risk of overfitting (\eg \prefixours may find one test case sufficient to rank) and of excessively aggressive tuning or non-regression assurance. Developers should balance these risks and gradually increase their confidence based on the computational budget.

\textbf{Cost saving potentials.}
A highlight of our results is, for example, defects4j, which can be reduced to only 5\% of the tests, saving up to hundreds of dollars per new LLM test.
These reductions can be observed on many different datasets at different scales. 
Our results are available in the supplementary material, and since our tool is available, this enables benchmark creators to easily check their benchmark.


\textbf{Conclusion.}
We have introduced the ranked test suite minimization (RTSM) and its solution. It is a new problem that aims to minimize a benchmark while keeping the performance ranking of the variants over the test suites intact. 
This problem is highly relevant for practitioners, as it enables cost-effective benchmarking for variant comparison, rapid regression detection, performance modeling, benchmark design, etc. 

Addressing this problem requires developing new approaches specifically designed to optimize benchmark selection while preserving variant ranking accuracy. We have introduced our approach \prefixours based on bisection sampling and supplemented with divide and conquer and iterative solving; and introduced baselines to compare to, including MILP, \prefix{PCA}, \prefix{RANDOM}, and others based on existing techniques.
Experiments on 50 different benchmarks show that divide-and-conquer enables a 4-times faster solving and that iterative solving always leads to improved results.
Iterative solving improves the solutions in 76\% of the cases where a cost reduction was found.
Our experiments show that \prefixours outperforms all alternatives.
Furthermore, when reducing the target Kendall, we show that our approach highly benefits from it, by reducing the Kendall by 1\% on average, we get a 10\% improvement.
We also showed that our method behaves well even if all variants are not present, it often introduces some imprecision; however, we showed that in most cases, this is an interesting trade-off, we gain more cost reduction than we lose Kendall or Spearman.

In general, with \prefixours, more than half of the benchmarks could be reduced with a mean reduction over these reducible benchmarks of 67\%.
This implies that quite extensive reductions are available for many benchmarks, including those used today.

\textbf{Future Work.}
Next steps in the direction of this work include studying the applications and implications of our methods in different domains.
For example, the field of AI is driven by benchmark, standardizing such minimization approaches could drastically reduce the cost of research comparison, thus increasing research velocity.
\new{An important direction is studying other formulations of the linear regression problem in order to make the divide and conquer algorithm always return solutions with the guaranteed target Kendall value.}
Another direction is to work with partial data, that is, with some variants running only on a subset of the tests.

\bibliography{references}

\newpage
\appendix

\section{Full Data}

{\centering 

    \begin{longtable}{@{}r|llllll@{}}
        \toprule \\ Fraction & \multicolumn{1}{l}{MILP} & \multicolumn{1}{l}{\prefixours} & \multicolumn{1}{l}{\prefix{PCA}} & \multicolumn{1}{l}{\prefix{random}} & \multicolumn{1}{l}{\prefix{greedy}}\\
\midrule \\
ASP-POTASSCO & 0.00 & 0.99 (0.02) & 0.27 (0.08) & 0.80 (0.27) & 0.23 (0.08)\\
BNSL-2016 & 1.00 & 1.00 & 0.53 (0.03) & 1.00 & 0.53 (0.05)\\
CSP-Minizinc-Obj-2016 & 0.94 & 0.61 (0.27) & 0.11 (0.13) & 0.22 (0.07) & 0.12 (0.06)\\
CSP-Minizinc-Time-2016 & 0.96 & 0.75 (0.27) & 0.11 (0.08) & 0.27 (0.08) & 0.13 (0.08)\\
CSP-MZN-2013 & 0.00 & 1.00 & 0.19 (0.21) & 0.94 (0.33) & 0.11 (0.11)\\
defects4j & 0.95 & 0.62 (0.52) & 0.00 & 0.10 (0.04) & 0.00\\
defects4j-cutoff-2024-10-01 & 0.98 & 0.92 (0.07) & 0.16 (0.04) & 0.99 & 0.09 (0.05)\\
defects4j-cutoff-2025-01-01 & 0.96 & 0.58 (0.52) & 0.02 (0.04) & 0.15 (0.04) & 0.01 (0.03)\\
gcc-ctime & 0.00 & 0.00 & 0.00 & 0.00 & 0.00\\
gitbugjava & 0.00 & 0.00 (0.01) & 0.00 & 0.00 & 0.00\\
gitbugjava-cutoff-2024-10-01 & 0.86 & 0.35 (0.14) & 0.01 (0.02) & 0.23 (0.15) & 0.00 (0.01)\\
gitbugjava-cutoff-2025-01-01 & 0.00 & 0.08 (0.09) & 0.00 & 0.01 (0.02) & 0.00 (0.01)\\
GLUHACK-2018 & 0.00 & 0.97 (0.07) & 0.35 (0.08) & 0.98 (0.08) & 0.21 (0.04)\\
humaneval-pass1 & 0.00 & 0.00 & 0.00 & 0.00 & 0.00\\
humaneval-plus-pass1 & 0.00 & 0.00 & 0.00 & 0.00 & 0.00\\
imagemagick-time & 0.00 & 0.03 (0.01) & 0.00 & 0.07 (0.07) & 0.00\\
imagemagick-time-size & 0.00 & 0.00 & 0.00 & 0.00 & 0.00\\
IPC2018 & 0.99 & 0.90 (0.18) & 0.26 (0.09) & 0.64 (0.15) & 0.18 (0.08)\\
lingeling-conflicts & 0.00 & 0.20 (0.02) & 0.09 (0.05) & 0.10 (0.04) & 0.05 (0.04)\\
lingeling-conflicts-reductions & 0.00 & 0.00 & 0.00 & 0.00 & 0.00\\
MAXSAT-PMS-2016 & 1.00 & 0.96 (0.07) & 0.36 (0.07) & 0.72 (0.20) & 0.29 (0.18)\\
MAXSAT-WPMS-2016 & 1.00 & 0.97 (0.04) & 0.34 (0.19) & 0.97 (0.05) & 0.26 (0.08)\\
MAXSAT15-PMS-INDU & 0.00 & 0.77 (0.23) & 0.17 (0.16) & 0.47 (0.05) & 0.13 (0.15)\\
nodejs-ops & 0.00 & 0.77 (0.32) & 0.11 (0.18) & 0.33 (0.19) & 0.06 (0.03)\\
OPENML-WEKA-2017 & 0.00 & 0.59 (0.47) & 0.00 & 0.17 (0.03) & 0.00\\
poppler-time & 0.00 & 1.00 & 0.37 (0.31) & 0.93 (0.21) & 0.38 (0.28)\\
poppler-time-size & 0.00 & 0.00 & 0.00 & 0.00 & 0.00\\
PROTEUS-2014 & 0.00 & 0.55 (0.07) & 0.08 (0.11) & 0.27 (0.14) & 0.07 (0.09)\\
QBF-2014 & 1.00 & 1.00 & 0.18 (0.15) & 0.93 (0.38) & 0.17 (0.11)\\
QBF-2016 & 0.00 & 0.63 (0.45) & 0.12 (0.14) & 0.31 (0.07) & 0.08 (0.11)\\
SAT03-16-INDU & 0.00 & 1.00 & 0.40 (0.28) & 0.94 (0.14) & 0.19 (0.15)\\
SAT11-INDU & 1.00 & 0.75 (0.37) & 0.20 (0.09) & 0.36 (0.06) & 0.14 (0.05)\\
SAT12-ALL & 0.00 & 0.24 (0.09) & 0.02 (0.02) & 0.17 (0.14) & 0.02 (0.06)\\
SAT15-INDU & 0.00 & 0.40 (0.24) & 0.10 (0.10) & 0.21 (0.05) & 0.09 (0.05)\\
SAT16-MAIN & 0.97 & 0.76 (0.11) & 0.25 (0.07) & 0.35 & 0.16 (0.07)\\
SAT18-EXP & 0.00 & 0.70 (0.28) & 0.10 (0.11) & 0.24 (0.05) & 0.14 (0.07)\\
SAT20-MAIN & 0.00 & 0.44 (0.15) & 0.11 (0.08) & 0.21 (0.08) & 0.09 (0.07)\\
sqlite-q1 & 0.00 & 0.17 (0.16) & 0.03 (0.06) & 0.07 & 0.02 (0.06)\\
sqlite-q1-q5 & 0.00 & 0.00 & 0.00 & 0.00 & 0.00\\
sqlite-q1-q5-q10 & 0.00 & 0.00 & 0.00 & 0.00 & 0.00\\
sqlite-q1-q5-q10-q15 & 0.00 & 0.00 & 0.00 & 0.00 & 0.00\\
TTP-2016 & 0.00 & 0.84 (0.20) & 0.05 (0.10) & 0.31 (0.30) & 0.03 (0.06)\\
x264-etime & 0.00 & 0.00 & 0.00 & 0.00 & 0.00\\
x264-etime-cpu & 0.00 & 0.00 & 0.00 & 0.00 & 0.00\\
x264-etime-cpu-size & 0.00 & 0.00 & 0.00 & 0.00 & 0.00\\
x264-etime-cpu-size-fps & 0.00 & 0.00 & 0.00 & 0.00 & 0.00\\
x264-etime-cpu-size-fps-kbs & 0.00 & 0.00 & 0.00 & 0.00 & 0.00\\
xz-time & 0.41 & 0.06 (0.13) & 0.00 & 0.28 (0.29) & 0.00\\
xz-time-size & 0.00 & 0.00 & 0.00 & 0.00 & 0.00\\
\bottomrule
    \caption{Mean Cost reduction of different methods with all variants with 95\% confidence interval in parenthesis if greater than 0.}\label{table:gain_full}
\end{longtable}
}
{\centering 

    \begin{longtable}{@{}r|llllll@{}}
        \toprule \\ Fraction & \multicolumn{1}{l}{MILP} & \multicolumn{1}{l}{\prefixours} & \multicolumn{1}{l}{\prefix{PCA}} & \multicolumn{1}{l}{\prefix{random}} & \multicolumn{1}{l}{\prefix{greedy}}\\
\midrule \\
ASP-POTASSCO & 0.89 (0.30) & 0.26 (0.10) & 0.76 (0.31)\\
BNSL-2016 & 1.00 & 0.43 (0.31) & 0.99 (0.02)\\
CSP-Minizinc-Obj-2016 & 0.67 (0.29) & 0.15 (0.17) & 0.23 (0.14)\\
CSP-Minizinc-Time-2016 & 0.63 (0.28) & 0.09 (0.09) & 0.24 (0.11)\\
CSP-MZN-2013 & 0.76 (0.71) & 0.14 (0.08) & 0.69 (0.50)\\
defects4j & 0.70 (0.57) & 0.01 (0.02) & 0.16 (0.19)\\
defects4j-cutoff-2024-10-01 & 0.74 (0.21) & 0.09 (0.13) & 0.51 (0.77)\\
defects4j-cutoff-2025-01-01 & 0.51 (0.43) & 0.02 (0.04) & 0.24 (0.26)\\
gcc-ctime & 0.12 (0.02) & 0.00 & 0.14 (0.07)\\
gitbugjava & 0.08 (0.13) & 0.00 & 0.02 (0.05)\\
gitbugjava-cutoff-2024-10-01 & 0.20 (0.25) & 0.00 (0.01) & 0.17 (0.21)\\
gitbugjava-cutoff-2025-01-01 & 0.14 (0.20) & 0.00 & 0.02 (0.04)\\
GLUHACK-2018 & 0.91 (0.20) & 0.21 (0.24) & 0.87 (0.41)\\
humaneval-pass1 & 0.00 & 0.00 & 0.00\\
humaneval-plus-pass1 & 0.00 & 0.00 & 0.00\\
imagemagick-time & 0.97 (0.01) & 0.05 (0.06) & 0.73 (0.21)\\
imagemagick-time-size & 0.00 & 0.00 & 0.00\\
IPC2018 & 0.82 (0.25) & 0.20 (0.15) & 0.46 (0.27)\\
lingeling-conflicts & 0.29 (0.06) & 0.07 (0.04) & 0.09 (0.06)\\
lingeling-conflicts-reductions & 0.00 & 0.00 & 0.00\\
MAXSAT-PMS-2016 & 0.91 (0.20) & 0.22 (0.27) & 0.65 (0.52)\\
MAXSAT-WPMS-2016 & 0.91 (0.14) & 0.30 (0.24) & 0.93 (0.33)\\
MAXSAT15-PMS-INDU & 0.97 (0.02) & 0.19 (0.26) & 0.58 (0.50)\\
nodejs-ops & 0.89 (0.54) & 0.09 (0.05) & 0.45 (0.44)\\
OPENML-WEKA-2017 & 0.77 (0.15) & 0.08 (0.11) & 0.28 (0.18)\\
poppler-time & 1.00 & 0.23 (0.29) & 0.84 (0.35)\\
poppler-time-size & 0.00 & 0.00 & 0.00\\
PROTEUS-2014 & 0.68 (0.23) & 0.06 (0.04) & 0.24 (0.17)\\
QBF-2014 & 0.98 (0.08) & 0.12 (0.08) & 0.60 (0.63)\\
QBF-2016 & 0.73 (0.42) & 0.10 (0.07) & 0.32 (0.20)\\
SAT03-16-INDU & 0.98 (0.08) & 0.25 (0.09) & 0.74 (0.45)\\
SAT11-INDU & 0.59 (0.23) & 0.13 (0.13) & 0.25 (0.15)\\
SAT12-ALL & 0.63 (0.40) & 0.04 (0.03) & 0.22 (0.20)\\
SAT15-INDU & 0.45 (0.27) & 0.10 (0.07) & 0.18 (0.10)\\
SAT16-MAIN & 0.75 (0.14) & 0.19 (0.15) & 0.32 (0.12)\\
SAT18-EXP & 0.70 (0.32) & 0.08 (0.06) & 0.24 (0.11)\\
SAT20-MAIN & 0.66 (0.26) & 0.07 (0.06) & 0.21 (0.10)\\
sqlite-q1 & 0.35 (0.10) & 0.09 (0.06) & 0.14 (0.09)\\
sqlite-q1-q5 & 0.00 & 0.00 & 0.00\\
sqlite-q1-q5-q10 & 0.00 & 0.00 & 0.00\\
sqlite-q1-q5-q10-q15 & 0.00 & 0.00 & 0.00\\
TTP-2016 & 1.00 & 0.08 (0.05) & 0.77 (0.50)\\
x264-etime & 0.68 (0.54) & 0.04 (0.03) & 0.46 (0.31)\\
x264-etime-cpu & 0.00 & 0.00 & 0.00\\
x264-etime-cpu-size & 0.00 & 0.00 & 0.00\\
x264-etime-cpu-size-fps & 0.00 & 0.00 & 0.00\\
x264-etime-cpu-size-fps-kbs & 0.00 & 0.00 & 0.00\\
xz-time & 0.17 (0.18) & 0.00 & 0.23 (0.23)\\
xz-time-size & 0.00 & 0.00 & 0.00\\
\bottomrule
    \caption{Mean Cost reduction of different methods with all variants with 95\% confidence interval in parenthesis if greater than 0.}\label{table:gain_99_full}
\end{longtable}
}

\end{document}